\documentclass[10pt,journal,compsoc]{IEEEtran_jnal}

\usepackage{graphics}
\usepackage{graphicx}
\usepackage{amsmath}
\usepackage{amsthm}
\usepackage{amsfonts}
\usepackage[linesnumbered,boxed]{algorithm2e}
\usepackage{cases}
\usepackage{xspace}
\usepackage{multirow}
\usepackage{slashbox}
\usepackage{subfigure}
\usepackage{listings}
\usepackage{color}
\usepackage{url}
\usepackage{ftnxtra}
\usepackage{pbox}
\usepackage{ragged2e}
\usepackage{indentfirst}
\usepackage[justification=centering]{caption}

\usepackage{algorithmicx}
\usepackage{algpseudocode}

\usepackage{amsmath}

\newcommand{\BRAID}{BRAID}

\newcommand{\tabincell}[2]{\begin{tabular}{@{}#1@{}}#2\end{tabular}}

\makeatletter
\renewcommand{\paragraph}{\@startsection{paragraph}{4}{\z@}
	{4pt plus 6pt}{\z@}{\normalsize\itshape}}
\makeatother

\ifCLASSINFOpdf
\else
\fi

\begin{document}

\title{Boosting API Recommendation \\ with Implicit Feedback}

\author{Yu Zhou, Xinying Yang, Taolue Chen, Zhiqiu Huang, Xiaoxing Ma, Harald Gall
\IEEEcompsocitemizethanks{\IEEEcompsocthanksitem Y. Zhou, X. Yang and Z. Huang are with College of Computer Science and Technology,
Nanjing University of Aeronautics and Astronautics, Nanjing, China. Y. Zhou is also with State Key Laboratory for Novel Software Technology, Nanjing University, Nanjing, China. \protect\\
E-mail:  \{zhouyu,xy\_yang,zqhuang\}@nuaa.edu.cn
\IEEEcompsocthanksitem T. Chen is with Department of Computer Science, University of Surrey, UK. He is also with State Key Laboratory for Novel Software Technology, Nanjing University, Nanjing, China. \protect\\
E-mail: taolue.chen@surrey.ac.uk
\IEEEcompsocthanksitem X. Ma is with State Key Laboratory for Novel Software Technology, Nanjing University, Nanjing, China.\protect\\
Email: xxm@ics.nju.edu.cn
\IEEEcompsocthanksitem H. Gall is with Department of Informatics, University of Zurich, Switzerland.\protect\\
E-mail: gall@ifi.uzh.ch
}
}


\IEEEtitleabstractindextext{
{\justifying
\begin{abstract}
Developers often need to use appropriate APIs to program efficiently, but it is usually a difficult task to identify the exact one they need from a vast list of candidates. To ease the burden, a multitude of API recommendation approaches have been proposed. However, most of the currently available API recommenders do not support the effective integration of user feedback into the recommendation loop. In this paper, we propose a framework, BRAID (\textbf{B}oosting \textbf{R}ecommend\textbf{A}tion with \textbf{I}mplicit Fee\textbf{D}back), which leverages learning-to-rank and active learning techniques to boost recommendation performance. By exploiting user feedback information, we train a learning-to-rank model to re-rank the recommendation results. In addition, we speed up the feedback learning process with active learning. Existing query-based API recommendation approaches can be plugged into BRAID.
We select three state-of-the-art API recommendation approaches as baselines to demonstrate the performance enhancement of BRAID measured by Hit@k (Top-k), MAP, and MRR. Empirical experiments show that, with acceptable overheads, the recommendation performance improves steadily and substantially with the increasing percentage of feedback data, comparing with the baselines.
\end{abstract}
}
}

\maketitle
\IEEEdisplaynontitleabstractindextext
\IEEEpeerreviewmaketitle

\IEEEraisesectionheading
{\section{Introduction}\label{introduction}}
 
\IEEEPARstart{A}{pplication} Programming Interfaces (APIs) play an important role in  software development \cite{brandt2009two}. With the help of APIs, developers can accomplish their programming tasks more efficiently \cite{Piccioni2013An}. However, due to the huge number of APIs in the library, it is impractical for developers to get familiar with all of them and always select the correct ones for specific development tasks.

To tackle this problem, many API recommendation approaches and tools have been proposed to relieve the burden of developers in understanding and searching APIs. Based on different inputs, there are generally two types of API recommendation scenarios, i.e., recommendation with queries and  recommendation without queries. \textcolor{black}{The first type requires} developers to state what is wanted \textcolor{black}{in natural language queries} which are fed into the recommendation system. For the second type, since there are no explicit queries, \textcolor{black}{the neighboring code snippets} will be leveraged as context, and the missing APIs will be inferred and recommended to end users. A majority of related work employs text similarity-based techniques. For example, some recommend APIs according to the similarity between search queries and supplementary information of APIs \cite{huang2018api, yu2016apibook}; some return API usages depending on how much they are related to context information in source code \cite{nguyen2019focus, fowkes2016parameter}. Generally, these approaches use keywords to narrow down the search scale in massive target repositories and speed up recommendation efficiency. However, in many cases, the correct API information is not literally similar to the query because of the semantic gap \cite{Haiduc2011On, Yang2012Inferring, li2018bridging}. For example, the answer to the query ``Make a negative number positive'' could be ``java.lang.Math.abs'', which returns the absolute value of the argument, matching the problem perfectly. For these dissimilar query-answer pairs, textual matching is of limited usage. Secondly, very few of these approaches consider the role of developers' feedback information in the recommendation process. Such information is usually crucial to improve the API recommendation performance.

Feedback information generally refers to \textcolor{black}{user} interaction information with the recommended results during a recommendation session. Usually, it reflects the user preference for different items. In traditional recommendation systems \cite{resnick1997recommender}, the use of feedback information could greatly improve the accuracy of recommendation \cite{salton1990improving, Carpineto2012A}. For example, in a movie recommendation system, the user viewing history is regarded as feedback information. In an online shopping system, feedback usually refers to the product browsing history of a particular customer. We note that they are usually referred to as implicit feedback. (In contrast, rating from users is considered to be explicit.)
\textcolor{black}{Implicit feedback indirectly reflects user opinion and could be collected by observing their behavior~\cite{hu2008collaborative}. The observable behavior may include selection, duration, repetition, purchase, etc.~\cite{oard1998implicit}}.
In the process of API recommendation, selecting an API from the recommended list usually suggests that the API is useful for the user to solve the particular problem specified in the query. Hence, it is deemed to be the correct answer to the query. During each query-answer session, we record the query alongside with the API selected by the user, inserting such a query-API pair into the feedback repository. The API is regarded as feedback information of the query, 
which can reveal, e.g., the user's programming habits. Moreover, in many cases, feedback from programmers actually provides answers to the queries and would play a significant role in processing similar user queries and improving the performance of the recommender in the future. This highlights the role of feedback in API recommendation systems, possibly in a more pronounced way than traditional recommendation systems. 
 
\textcolor{black}{When searching information via browsers, people usually pay attention to the first few results returned by the search engine. Likewise, ideally the API which match the user query should  be put on the top of the list.
From the user's perspective, when they are not familiar with any of the APIs on the recommendation list, they are more likely to pick up the top-ranked API. 
}
In this paper, we propose a novel framework, BRAID (\textbf{B}oosting \textbf{R}ecommend\textbf{A}tion with \textbf{I}mplicit Fee\textbf{D}back), to  boost recommendation effectiveness by leveraging (implicit) feedback information. Particularly, we focus on the first type of recommendation scenarios, i.e., recommendation with queries. By introducing feedback, not only do we improve the performance of API recommendations, but also we can accomplish personalized recommendation. For the same query, \textcolor{black}{different list orders could be recommended based on each user's} personal interaction history (i.e., feedback). 
Moreover, our framework could accommodate existing recommendation approaches as components.

To effectively integrate user feedback into the code recommendation loop, we harness learning-to-rank (LTR) techniques, which are widely used in areas such as information retrieval and recommendation.
The key of LTR in information retrieval is to train a ranking model by which a given query can decide an optimized order of the relevant documents based on feedback information. 
By viewing APIs as documents, we can apply LTR techniques to API recommendation to boost its performance. \textcolor{black}{In particular, we leverage \emph{related information features} and \emph{feedback features} to train the model (cf.\ Section 3.2). The former  consists of API path features and API description features, representing the relevance of the recommended APIs and the associated document descriptions respectively; the latter represents the relevance to the APIs in the feedback repository.} 
Furthermore, to accelerate the feedback learning process, we incorporate active learning which is to alleviate the ``cold start" 
of tenuous feedback information at the beginning.
\textcolor{black}{We collect query-API pairs by leveraging crowdsourced knowledge, which function as an oracle to provide the correct label. These pairs are then put to the training set.} By iterating this process we can obtain a well-trained active learning model with the expanded labeled set. This training set can be, in turn, used to train a \textcolor{black}{well-performed model} to generate an optimized recommendation list.

To demonstrate the effectiveness of BRAID, we select three recent state-of-the-art API recommendation systems, i.e., BIKER~\cite{huang2018api}, RACK~\cite{Rahman2016RACK,rahman2019automatic}, NLP2API~\cite{rahman2018effective}, as baselines and Hit@k/Top-k accuracy, MAP, MRR as evaluation metrics. 
With continuous accumulation of feedback information, the Top-$1$ accuracy is increased by 9.44\%, 6.79\%,
18\% and 18.39\% for BIKER (method level), BIKER (class level), RACK and NLP2API respectively.

The main contributions of the paper are as below.

\begin{itemize}
\item We propose a novel framework \BRAID\footnote{\url{https://github.com/yyyxy/vscode-plugin-for-braid/}}, which integrates programmers' feedback information by using the learning-to-rank technique to improve the accuracy of API recommendation.
\item BRAID also features the active learning technique, with which the learning process of feedback information can be accelerated. Even with a small proportion of feedback data, the performance of recommendation can still be enhanced considerably.
\item We conduct a comprehensive empirical study and compare BRAID to three state-of-the-art API recommendation systems. The results show that our approach performs well and \textcolor{black}{demonstrate its generalizability}.

\end{itemize}

Our work is orthogonal to the recent efforts in recommending APIs with machine learning techniques, largely in the context of intelligent software development. It is not to put forward yet another recommendation method, but is to boost the performance 
and is applicable to a wide spectrum of \textcolor{black}{existent} query-based recommendation systems. To the best of our knowledge, 
\textcolor{black}{this represents one of the first works to combine LTR, active learning and feedback information in API recommendation.}

\medskip
\noindent\emph{Structure of the paper.} 
Section \ref{background} briefly introduces the background of this study. Section \ref{approach} gives the details of our approach. Section \ref{experiment} presents the experimental settings and comparative results on related API recommendation systems. In section \ref{threats} and \ref{related}, threats to validity and related work are discussed respectively. Finally,  conclusion is drawn and future research is outlined in Section \ref{conclusion}.

\section{Background}\label{background}

\subsection{Learning-to-rank}\label{bg:LTR}
As a widely used ranking technique, LTR has achieved great success in a variety of areas including information retrieval, natural language processing, and software engineering \cite{liu2009learning, li2011learning,ye2014}. The basic task of LTR is to learn $k$ ordered documents $d=(d_1, \cdots d_k)$ from the document set $D$ by optimizing a loss function which is dependent on a given query $q$. LTR is essentially a supervised learning task, typically by extracting features from documents and predicting the corresponding labels which reflect the relevance between the query and the documents. Different from traditional approaches based on similarity calculation, the main characteristic of LTR is to define a loss function and train a ranking model $f(q, d)$ 
to sort the candidate documents in $d$. In this work, in a nutshell, we regard APIs as ``documents", and cast API recommendation as an LTR problem.  

LTR techniques can be classified based on the underlying learning model. Examples include SVM techniques~\cite{cao2006}, boosting techniques~\cite{freund03}, neural network techniques~\cite{song2014}, and others~\cite{li2011learning}. A more interesting classification is based on the characteristics of the input space, where one usually speaks of pointwise, pairwise and listwise LTR \cite{xia2008listwise,cao2007learning}.
In general, the pointwise approach focuses on the relevance of a query and a single document. By converting each single document into a feature vector, it can predict the relevant score of the document via classification or regression methods. 
The pairwise approach regards ranking as comparing the relative preference between document pairs. In this way, it turns a ranking task into deciding the relative order of each document pair, which can be considered as a binary classification or a pairwise regression problem. 
The listwise approach takes the results of the user query (namely, a list of documents) as a data point in the training data set  based on which a ranking model $M$ can be trained. For a new query, $M$ predicts each document on the list for the new query and then ranks them in (say) descending order.

In API recommendation, it is neither practical nor necessary to obtain a fully ranked list of APIs, since programmers are merely interested in the most appropriate APIs associated with the query and ignore the irrelevant ones. Instead we only need to compare pairwise preference of a few candidate APIs 
with the help of programmers' feedback. 
\textcolor{black}{On the other hand, in general pairwise approaches work better in practice than pointwise approaches because predicting relative order is closer to the nature of ranking than predicting class labels or relevance scores.}
As a result, in our framework, we adopt the pairwise LTR technique.

\subsection{Active Learning}\label{bg:AL}
Supervised learning requires annotated/labeled data, which may be very expensive to obtain in many cases. Active learning is proposed with the general aim to train a model of better performance but with fewer training instances. When the annotated data is scarce or the cost of labeling data is high, the active learning algorithm can actively select specific data to label; these data will then be sent to annotators. Generally speaking, the selected samples should be the most informative ones, which can not only make a maximum contribution to model optimization, but also help reduce the amount of annotated data \cite{settles2009active}.

Generally speaking, the paradigm of active learning can be represented as a tuple $A=(C, S, O, F, U)$, where $C$ is the model to be learnt (e.g., a classifier), $S$ denotes the query function which acquires the most informative data from unlabeled samples, and $O$ represents the oracle which labels the samples. In addition, $F$ and $U$ are the sets of labeled and unlabeled samples respectively.

An active learning algorithm usually starts by training a model with only a small amount of labeled data from $F$. Then it inquires the function $S$ which defines the selection strategy, and thus obtains the samples from the unlabeled data set $U$. As the next step, it submits these selected samples to the oracle $O$ for annotation and inserts them into the labeled set when they are returned. Finally, the newly labeled samples are used to retrain the model. This process repeats until some specific termination criteria  are met, such as those based on the number of iterations or performance related metrics.

\section{Approach}\label{approach}

\begin{figure}[!h]
	\centering
	\includegraphics[height=4cm, width=8.9cm]{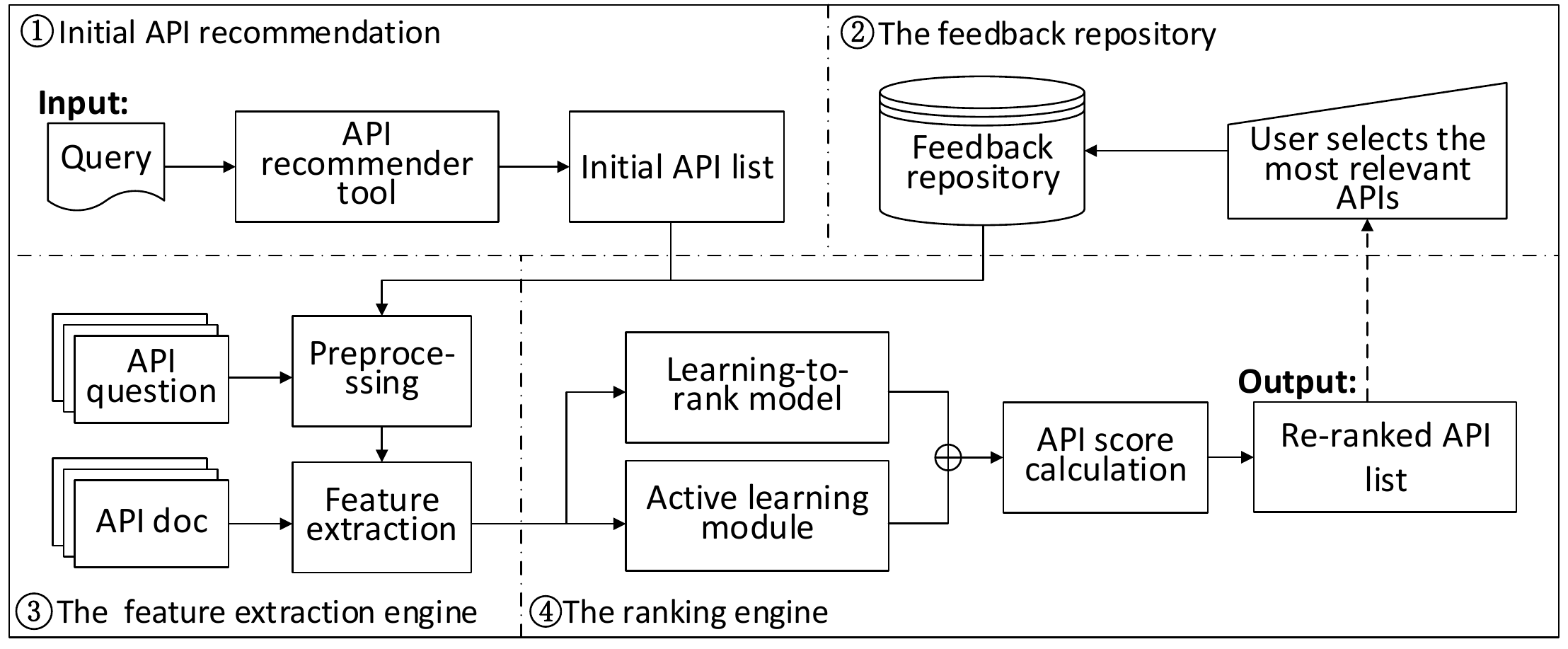}
	\caption{The overview of BRAID\label{fig4}}
\end{figure}

As illustrated by Fig.~\ref{fig4}, the BRAID framework mainly consists  of four parts.
\begin{enumerate}
\item[(a)]  \emph{Initial API recommendation}. Given a query as input, an initial API recommendation list is returned. This could be acquired by applying the existing API recommendation algorithms to the given query.

\item[(b)] The \emph{feedback repository} which stores pairs of queries and associated recommended APIs. More formally, the feedback repository $FR$ is a set of pairs $(Qu, Ap)$ where $Qu$ is a query and $Ap$ is the corresponding APIs.  \textcolor{black}{When a user selects certain APIs from a recommendation list, the observable behavior will be tracked, i.e., the query and the selected APIs are to be recorded in the feedback repository.} Initially, the feedback repository is empty, but will accumulate in the course of interactions with users. 
	
\item[(c)] The \emph{feature extraction engine} which generates a feature vector for each API on the recommended API list when a query is given. The feature vector comprises two parts, i.e., \emph{feedback features} and \emph{related information features}. In particular, the feedback information is obtained by looking up the feedback repository whereas the related information is obtained from relevant domain knowledge, e.g., Java official API document information (cf.\ Section \ref{sec:feature}).

	
\item[(d)] The \emph{ranking engine} which ranks the recommended APIs for a given query. To this end, the engine applies two techniques: (1) LTR to compute scores based on the generated feature vectors (cf.\ Section \ref{sec:LTR}); and (2) active learning which leverages crowdsourced knowledge (from, e.g., Stack Overflow) as an oracle and trains a classifier to predict the score (cf.\ Section \ref{sec:AL}). The two scores are combined to give the final verdict (cf.\ Section~\ref{sec:reorder}).
	
\end{enumerate}

The basic workflow of our approach is as follows.
\begin{enumerate}
\item When a user makes a query $Q$ to the system (in the form of, for instance, a short sentence in \textcolor{black}{a} natural language), a base API recommendation method 
is employed to provide an initial API list $L_Q$. 
	
\item The system looks up the feedback repository $FR$, checking whether or not there is a query similar to the user query $Q$. If this is the case, the system returns a set $SP$ of query-API pairs where  the similarity score of each query with $Q$ is above a certain threshold $\epsilon$ (cf. Section~\ref{sec:similarity}), i.e.,
\begin{align*}
SP:=\{(Qu, Ap) & \mid (Qu, Ap)\in FR \\
                      & \mbox{ and } sim(Qu, Q) \geq \epsilon \}
\end{align*}
Otherwise, there is no available query in  $FR$ similar to $Q$ (which is especially the case at the initial stage of the interaction), and $SP$ is simply an empty set.
The recommended APIs in $L_Q$ and $SP$ are to be fed to the feature extraction engine.  
	
\item The feature extraction engine, upon receiving $L_Q$ and $SP$, computes a composite feature vector $FV$. $FV$ includes two components, i.e., $FF$ and $RIF$. The former corresponds to the feedback features, while the latter corresponds to the related information features. (In case that $SP$ is empty, \textcolor{black}{$FV$ consists} solely of related information features.)
	
	\item The ranking engine takes 
	$FV$ as input, and applies the trained learning-to-rank model and active learning model to obtain the prediction values. The system then calculates the API scores based on the prediction values of these two models. Afterwards $L_Q$ is re-ranked in descending order according to the API scores, and new recommendations are presented to the users.
\end{enumerate}


%
%

As the core component of our framework, the feedback repository is maintained throughout the life of the system and is \textcolor{black}{kept up-to-date} with the interaction of the users. In the beginning, the feedback repository is empty. (In this case, no feedback feature can be provided, and thus BRAID outputs the initial API recommendation list as a result.)  
When the APIs are recommended to the users (e.g., programmers) who are supposed to implicitly label the most relevant APIs which are treated as the ``ground-truth" recommendation of the given query,  the query-API pair would be the feedback from the user and is stored in the feedback repository. The feedback repository grows gradually along with more user interactions. 

In general, the feedback repository is used in both feature extraction 
and training the LTR model (cf.\ Section~\ref{sec:LTR}). We note that, for efficiency consideration, we do not re-train the LTR model  every time the feedback repository is updated. Instead it is done on a user session basis, \textcolor{black}{which in our context denotes a series of interactions continuously performed by a specific user, for instance, when the user launches the API recommender followed by a number of queries.}
%
In this way we can strike a balance between ranking precision and overheads.
%


\subsection{Preprocessing and similarity calculation} \label{sec:similarity}

To facilitate feature extraction and learning steps, we first need to convert user queries and APIs (as well as their related documents) into vectors. 
As mentioned in Section~\ref{introduction}, the lexical 
gap between queries in natural languages and APIs in programming languages impedes the recommendation performance. We hence use word embedding to bridge such a gap during vectorization. To train the model, we collect API related posts in Stack Overflow website.\footnote{https://stackoverflow.com/} Particularly we use the data dumped from Stack Exchange.\footnote{https://archive.org/download/stackexchange/stackoverflow.com-Posts.7z, updated in March 2019} 
All the titles of the posts which are tagged with Java are extracted in particular, since we mainly focus on Java related API recommendation. (Note however that the general methodology is clearly not Java-specific.) The remaining posts are subject to classic textual preprocessing steps including tokenization and stemming. NLTK\footnote{http://www.nltk.org/} is employed to fulfil the pre-processing task, and Word2Vec\footnote{https://radimrehurek.com/gensim/models/word2vec.html} is used to train the embedding model. \textcolor{black}{Similar to Huang et al.~\cite{huang2018api}}, we calculate the IDF (Inverse Document Frequency) of each word in the preprocessed post corpus, and thus build an IDF vocabulary as the weighting schema of the embedding model.



\medskip
\noindent\textbf{Similarity calculation}.
%
To calculate the similarity between a user query $Q$ and a text $S$ (e.g., the query stored in the feedback repository), we first convert them to two bag-of-words $Q$ and $S$. Then we use the semantic similarity measure introduced by Mihalcea et al.~\cite{mihalcea2006corpus}.

For any $w\in Q$, $sim(w,S)$ is defined to be  the maximum value of $sim(w,w')$ for each word $w'\in S$. Formally
\begin{equation}\label{eq:1}
sim(w,S)= \max \limits_{w' \in S} sim(w,w')
\end{equation}
where $sim(w,w')$ is the semantic similarity of the two words $w$ and $w'$, captured by the cosine distance of the embeddings of $w$ and $w'$ as vectors:
\begin{equation}\label{eq:2}
sim(w,w')= \frac{\vec{V}_{w} \cdot \vec{V}_{w'}}{\left|\vec{V}_{w}\right|\left|\vec{V}_{w'}\right|}
\end{equation}

Based on Equation~\eqref{eq:1}, the asymmetry similarity can be defined as:
\begin{equation}\label{eq:3}
sim^{a}(Q,S)= \frac{\sum_{w \in Q}sim(w,S)*idf(w)}{\sum_{w \in Q} idf(w)}
\end{equation}
where $idf(w)$ is computed as the number of documents that contain $w$.

Finally, the (symmetric) similarity between $Q$ and $S$ is derived by the arithmetic mean of $sim^a(Q,S)$ and $sim^a(S,Q)$, i.e.,
\begin{equation}
sim(Q, S)= \frac{sim^a(Q,S)+sim^a(S,Q)}{2}
\end{equation}

In this way, we can compute the similarity between user query and other artifacts such as  API, query in feedback repository, etc.
Recall that in step 2), the system needs to check whether there exists a query in the feedback repository which is similar to the user query. For this purpose, we set a parameter $\epsilon$ as the similarity threshold to distinguish whether or not two queries $Q$ and $S$ are similar. If  $sim(Q, S)\geq \epsilon$, then they are considered to be relevant. (Our experiment, via trial-and-error, empirically indicated that $\epsilon=0.64$ is a suitable configuration.)


\subsection{Feature extraction} \label{sec:feature}

Recall that the basic functionality of the feature extraction module is to compute the features of APIs. As stated in \emph{workflow 3)}, the input of this module is $SP$ and $L_Q$, where $L_Q$ is the recommended top-$N$ APIs for the query $Q$ and $SP$ is a set of query-API pairs stored in the feedback repository which crucially, corresponds to queries similar to $Q$. The aim is to generate a feature vector for each of the $N$ APIs in $L_Q$, based on $SP$. As the feature extraction is based on the query $Q$, this can be treated as a process of \emph{query-aware} feature engineering.

The rationale is that the relevance of each API in the recommended API list $L_Q$ to the user query $Q$ depends on (1) the relevance of the API-related description information to $Q$, and (2) whether in the feedback repository some API exists for dealing with a similar query. As a result, we consider
\begin{itemize}
\item \emph{related information features}, representing the 
relevance to the recommended APIs as well as the associated document description; 
\item \emph{feedback features}, representing the 
relevance to the APIs in the feedback repository. 
\end{itemize}
which are articulated as follows.

\smallskip
\noindent\textbf{Related information feature.}
The related information feature of each API on the recommended API list consists of the following two parts.	
\begin{itemize}
\item[(1)] API \textcolor{black}{path feature, representing the similarity between the user query $Q$ and the API path information.}
\textcolor{black}{Here an API path is represented by package name and class name, which is taken from the original API recommender tool} under consideration.  
	\item[(2)] API description feature, representing the similarity between the description under consideration and the user query $Q$. 
	The description can be obtained via  official API documentation. Particularly, we extract \textcolor{black}{the summary sentence describing the API class/method out of the official JDK 8 documentation.}
\end{itemize}
\textcolor{black}{In both cases, the similarity measure is calculated by the approach in Section \ref{sec:similarity}.}

\textcolor{black}{
\smallskip
\noindent\textbf{Example}. As an example, 
we consider the query $Q$ ``killing a running thread in Java" and the API $m$ 
on the top of the list in Table~\ref{returnlist}. Note that $m$ is from the package 'java.lang.Thread.start'. The API path feature of $m$ is the similarity between $Q$ and $m$.
The API description of $m$ is ``Causes this thread to begin execution; the Java Virtual Machine calls the run method of this thread", so the API description feature is set to be the similarity between $Q$  and the description both of which are treated as bags of words.
}

\smallskip
\noindent \textbf{Feedback feature.}
Feedback feature is extracted based on the similarity between a user query $Q$ and queries in feedback repository $FR$.




Recall that
\[SP:=\{(Qu, Ap)\mid (Qu, Ap)\in FR \mbox{ and }  sim(Qu, Q) \geq \epsilon\}\]

We then collect a subset of $SP$ consisting of only those whose API appears in $L_Q$, namely, $ST$.
%
Formally, we define $ST$ as below.
\begin{align*}
ST:=\{ & (Qu, Ap, sim(Q, Qu))  \mid \\
												& \qquad  (Qu, Ap)\in SP\mbox{ and } Ap\in L_Q\}
\end{align*}

We remark that there may be several tuples in $ST$ whose $Ap$ is the same. Therefore, an API in $L_Q$ may have several similarities to be considered as the feedback feature, and we select the most relevant five as the feedback feature.


\SetAlFnt{\footnotesize}
\begin{algorithm}
	\KwData{$ST$: tuple set, $FR$: feedback repository, and $L_Q$: initial API list}
	\KwResult{$FF$: Hashmap of feedback feature of the APIs in $L_Q$}
	\Begin{
		\tcc{Initialize $FF$ for API entries in $L_Q$;}
		$FF \longleftarrow new\ Hashmap()$\;
		\tcc{sort $ST$ in descending order based on the similarity score;}
		$ST \longleftarrow \textit{sortedBySim(ST)}$\;
	    \ForEach{API $\in L_Q$}{$index \longleftarrow 0$\;
        $ff \longleftarrow new\ Array[5]$;
	    	
	    \ForEach{$st \in ST$}{\uIf{$st.Ap == API$}{$ff[index] \longleftarrow st.sim(Q,Qu)$;}\uElse{$ff[index] \longleftarrow 0;$}
	    	
	    \uIf{$index < 5 $}{$index \longleftarrow index + 1;$}\uElse{break;}}

    	\tcc{add API and feature value pair into feedback vector $FF;$}
        $FF.put(API,ff);$}
    } \caption{Algorithm for generating feedback features}\label{alg:FF}
\end{algorithm}

Algorithm~\ref{alg:FF} shows the pseudo-code of feedback feature generation for $L_Q$. We first create an object $FF$ of \emph{Hashmap} type to accommodate the result (Line 2); then we sort $ST$ in descending order based on the similarity score (Line 3). Afterwards, we iterate the $L_Q$, and for each $API$, we create an array $ff$ (Line 6) to record the most relevant 5 similarity values with the API, from the sorted $ST$ (Line 7-16). Then, the $API$ and $ff$ pair is inserted into $FF$ (Line 19). 

\smallskip
\noindent\textbf{Example}.
\textcolor{black}{
	To continue with the previous example,} we firstly obtain the recommended API list $L_Q$ shown in Table~\ref{returnlist} from an initial API recommendation tool (e.g., BIKER),  and the $RIF$ of $L_Q$. Then we look up the feedback repository $FR$, finding a pair $SP(Qu, Ap)$ whose query is similar to $Q$ shown in Table \ref{q-a}. Because $Ap$ 'java.lang.Thread.interrupt' of the $SP$ is in $L_Q$ (the ninth API), this $SP$ and the similarity between $Qu$ and $Q$ can make up the tuple $ST$. The similarity is calculated as 0.72 based on the Equation (4). There is \textcolor{black}{no other $ST$}, so we put the similarity (0.72) into the first position of the feature vector $ff$, and the rest four elements would be zero.  $ff$ and $Ap$ form $FF$. Combining $FF$ with the $RIF$ together forms feature vectors $FV=(FF, RIF)$ of the APIs in the $L_Q$.

\begin{table}[!t]
	\centering
	\caption{The recommended API list of the query\label{returnlist}}
	\begin{tabular}{|c|p{0.1cm}p{6.2cm}|}
		\hline
		Query & \multicolumn{2}{|c|} {killing a running thread in java} \\
		\hline
		\multirow{10}{1.2cm}{Initial API list (by BIKER \cite{huang2018api})} & 1 & java.lang.Thread.start   \\
		\cline{2-3}
		&2 &  java.lang.Thread.stop \\
		\cline{2-3}
		&3  & java.lang.Thread.join   \\
		\cline{2-3}
		&4  & java.util.concurrent.Executor.newFixedThreadPool   \\
		\cline{2-3}
		&5   & java.lang.Process.destroy   \\
		\cline{2-3}
		&6  & java.lang.Thread.currentThread   \\
		\cline{2-3}
		&7  & java.lang.Thread.isAlive   \\
		\cline{2-3}
		&8    & java.util.concurrent.Executor.execute   \\
		\cline{2-3}
		&9  & java.lang.Thread.interrupt   \\
		\cline{2-3}
		&10  & java.lang.Object.wait   \\
		\hline
	\end{tabular}
\end{table}

\begin{table}[!t]
	\centering
	\caption{The similar query in the feedback repository\label{q-a}}
	\begin{tabular}{|c|c|}
		\hline
		Query & Stopping looping thread in Java \\
		\hline
		Answer & java.lang.Thread.interrupt \\
		\hline
	\end{tabular}
\end{table}


\subsection{Re-ranking recommendation API list} \label{sec:re-rank}

In this section, we describe the functionality of the ranking engine. As stated earlier, the input is a list of APIs produced by the adopted recommendation tool, endowed with feature vectors based on the user query. The ranking engine aims to re-rank the APIs on the list so the recommendation is more customised to the user feedback. To this end, we harness two techniques, i.e., LTR and active learning.
\textcolor{black}{In this framework, the LTR and active learning modules are independent. The active learning algorithm determines whether the unlabeled data is relevant. With the oracle, the labeled dataset 
is expanded, which can be used to improve the classifier in the active learning. 
The labeled dataset is used as the training set for LTR. Both LTR and active learning models predict  the API relevance scores for the given query, which are to be integrated as per Equation~\eqref{equa:sum_pred}.}



\subsubsection{LTR model and rank scores} \label{sec:LTR}







LTR is a supervised learning approach, which demands labeled training data. To this end, we use the recommended APIs for the queries 
stored in the \textcolor{black}{feedback repository}. Recall that each query-API pair $(Qu, Ap)$ in the feedback repository has gone through \textcolor{black}{feature} engineering (Section \ref{sec:feature}). 
We can then collect the feature vectors of the APIs in $L_{Qu}$, and label the selected API (i.e., $Ap$) as $1$, and others as $0$.
This process 
gives rise to the labeled training data set for the LTR model. 

We adopt LambdaMart~\cite{burges2010from}, a widely-used algorithm for ranking, as our LTR model.
LambdaMART is a boosted tree model with the optimization strategy based on LambdaRank~\cite{DBLP:conf/nips/BurgesRL06}. The key observation of the optimization strategy is that, in order to train a model, only the gradient of the objective function is needed, 
which can be modeled by the sorted positions of the items for a given query. In LambdaMART, we assume that there is an implicit objective utility function \emph{Util} whereby we define
\begin{equation}\label{equa:lambda1}
 \lambda_{ij} = \frac{\partial Util(s_i-s_j)}{\partial s_i} =\frac{-\sigma \left|\Delta Z_{ij}\right|}{1+ e^{-\sigma(s_{i}-s_{j})}}
\end{equation}
where for two feature vectors $V_{i}$ and $V_{j}$ such that $V_{i}$ ranks higher than $V_{j}$, $s_{i}$ and $s_{j}$ represent the scores of $V_{i}$ and $V_{j}$ respectively. $\sigma$ is a parameter of the sigmoid function the value of which determines the shape of the function. $\Delta Z_{ij}$ is the difference of a specific ranking metric calculated by swapping the rank positions of $V_{i}$ and $V_{j}$. For example, when $|\Delta Z_{ij}|$ stands for the change of metric MAP, such a model actually optimizes MAP directly.

Symmetrically, in case that $V_j$ ranks higher than $V_i$, we define
 \begin{equation}\label{equa:lambda2}
 \lambda_{ij} = \frac{\sigma \left|\Delta Z_{ij}\right|}{1+ e^{-\sigma(s_{i}-s_{j})}}
\end{equation}



With Equation~\eqref{equa:lambda1} and Equation~\eqref{equa:lambda2}, the gradient of $Util$ with respect to a feature vector $V_{i}$ can be written as:
\begin{equation}\label{equa:lambda3}
\lambda_{i} = \sum_{j\neq i} \mathbb{I}(i,j)\lambda_{ij}=\mathbb{I}(i,j)\frac{\sigma \left|\Delta Z_{ij}\right|}{1+ e^{-\sigma(s_{i}-s_{j})}}
\end{equation}
where $\mathbb{I}$ is the indicator function defined as:
\[\mathbb{I}(i,j)=\left\{\begin{array}{ll}
-1, &\text{if $V_i$ ranks higher than $V_j$},\\
~1, &\text{if $V_i$ ranks lower than $V_j$}.
\end{array}\right.\]

It follows that, for each feature vector $V_{i}$ ,
we can define the utility function as 
\begin{equation}\label{equa:loss}
Util_i = \sum_{j\neq i }  \left|\Delta Z_{ij}\right|\log(1+ e^{-\sigma(s_{i}-s_{j})})
\end{equation}

Since we build the LTR model based on  the tree-based algorithms \cite{chen2016xgboost}, the regularization term is based on the complexity of the tree model. More concretely, it is defined as
\begin{equation}
\Omega =\gamma T+\frac{1}{2}\beta \sum_{j=1}^{T} \left|\left|\omega_{j}\right|\right|^{2}
\end{equation}
where $T$ represents the number of the leaf node, $\omega$ is the weight of the leaf node, $\gamma$ and $\beta$ are hyper parameters used to adjust the weights of $T$ and $\omega$. (The experimental results show that $\gamma$ is set to 0.3 and $\beta$ to 1 in our setting.)

Finally, the objective of our LTR model is to maximize
\begin{equation}
\sum_i Util_i - \Omega.
\end{equation}
where 
$i$ ranges over all labeled samples.

LambdaMART trains a boosted tree model MART (multiple additive regression trees), in which the prediction value of the model is a linear combination of the outputs of a set of regression trees. In our LTR model, the LambdaMART maps the feature vector $V \in \mathbb{R}^d$ to $Score(V) \in \mathbb{R}$, which can be written as:
\begin{equation}\label{equa:score}
  Score(V)= \sum_{j = 1}^N \alpha_jf_j(V)
\end{equation}
where $f_j: \mathbb{R}^d \rightarrow \mathbb{R}$ is a function modeled by a single regression tree and the $\alpha_j \in \mathbb{R}$ is the weight associated with the $j$-th regression tree. Both  $f_j$ and $\alpha_j$ are learned during training, and $N$ is the number of trees.

For the given user query $Q$, we extract features as in Section \ref{sec:feature}, and then use the trained LTR model to predict the rank score for the recommended API list $L_{Q}$. The result is denoted by $Score_{Q}$, which comprises $Score(V)$ for all feature vectors $V$ of each API in $L_{Q}$.
\subsubsection{Active learning model and relevance scores} \label{sec:AL}

\begin{figure}[!h]
    \centering
    \includegraphics[height=3.2cm, width=9cm]{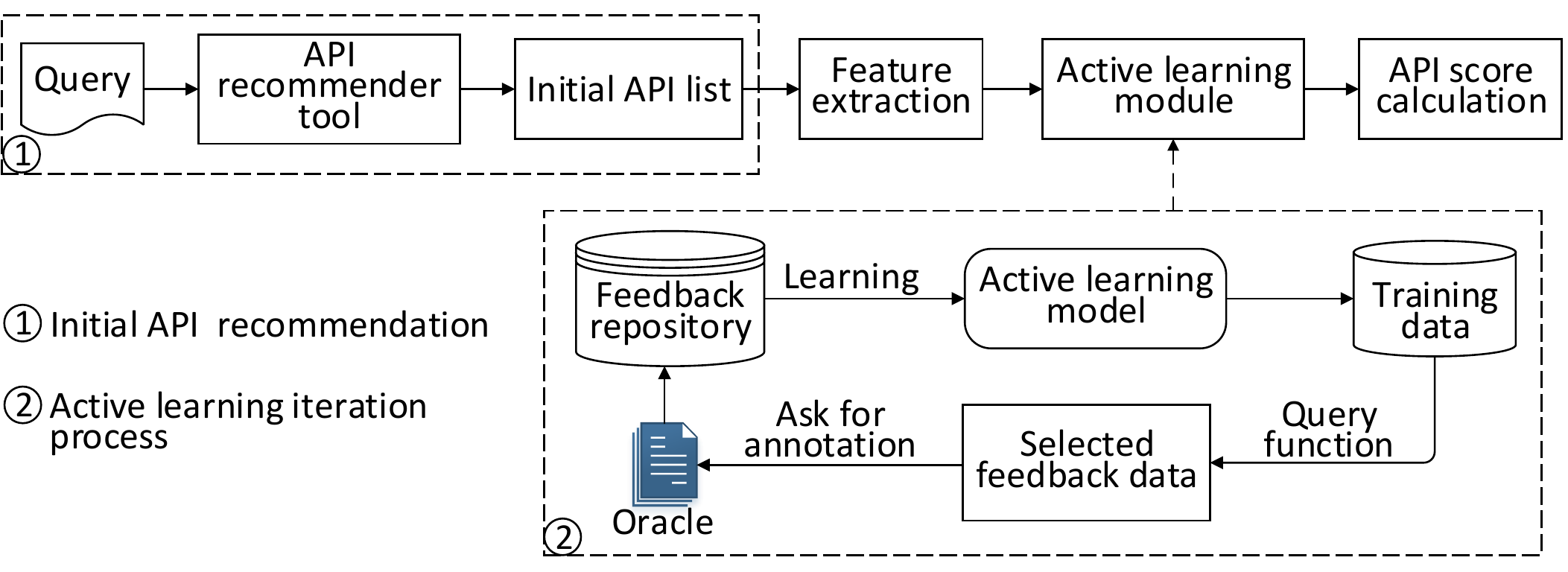}
    \caption{Active learning module architecture\label{fig5}}
\end{figure}

We utilize the active learning technique to improve the learning efficiency when the feedback repository data is scarce. An active learning algorithm usually starts by training a model with selective labeled data 
for which we follow the same approach as  LTR (cf.\ Section \ref{sec:LTR}). 
The structure of the active learning module is shown in Fig.~\ref{fig5}. 

For the active learning paradigm $A=(C, S, O, F, U)$, we use the Logistic Regression algorithm to train a model $C$. The uncertainty sampling strategy \cite{Lewis1994Heterogeneous} is used to select the most informative data (which may not be classified well by the classifier) as the query function $S$. 
\textcolor{black}{Specifically, we use a general framework of uncertainty sampling strategy, viz.\ least confidence $LC$~\cite{CulottaM05}, to select sample with the highest uncertainty value. The uncertainty value for the sample is defined as follow:
	\begin{equation}\label{equa:uncertainty}
	x^{*} = \arg\max\limits_{x} 1 - P_C(\hat{y}|x)
	\end{equation}
	where $\hat{y}$ is the class label with the highest posterior probability for the sample $x$ under the classifier $C$. }
In our work, we collect the query-API pairs to serve as the oracle $O$. 
These query-API pairs represent crowdsourced knowledge derived from the questions and accepted answers in Stack Overflow posts, which can be used to annotate the selected data. 
\textcolor{black}{We manually examine the dataset to assure its quality (cf.\ Section 4.1).
} Note that this is just one way to instantiate the oracle; one can certainly seek other resources to serve as the oracle. 

Because BRAID outputs the initial API recommendation list when the feedback repository is empty, the active learning module commences to play its role when the feedback data is available. 
		
			
		
		

First, we collect the feature vectors of the APIs in $L_{Qu}$ (cf. Section \ref{sec:LTR}) and label them to form the labeled set $F$. We formulate as a classification problem, and accordingly, use $F$ to build an active learning classifier model $C$. 
Next, we collect the feature vectors of the recommended APIs of the queries \textcolor{black}{whose topic is similar with the given query} in Stack Overflow to form the unlabeled set $U$. After applying the current model $C$ to the unlabeled set $U$, we use the uncertainty sampling strategy $S$ on $U$ to select data for which the classifier $C$ is less certain. 
 
%
Then the queries based on the selected data are sent to the oracle $O$ for annotation, and the results 
will be put into the feedback repository.
The selected samples will be used for expanding the labeled set along with their labels to retrain the classifier model $C$. 
The above steps are repeated, and we finally obtain an optimized classifier and an expanded feedback repository which will also be  used to train the LTR model (cf.\ Section \ref{sec:LTR}).

Similar to LTR, we consider the features extracted from a given query (cf.~Section \ref{sec:feature}) as input, and use the well-trained classifier to predict the relevance  of each API on the recommended list, where 
the relevance score simply takes the probability returned by the classifier. $Relev_{Q}$ is then 
obtained by computing the relevance score for the recommended API list of a user query $Q$. In this way, we can combine active learning with API recommendation systems.

\subsubsection{Re-ranking list and collecting user feedback}\label{sec:reorder}
The last step is to re-rank the API list. 
In Section \ref{sec:LTR} and Section \ref{sec:AL}, we have obtained the predictions of the API ($Score_{Q}$ and $Relev_{Q}$) of the $L_Q$ through well-trained LTR and active learning models respectively. By normalizing $Score_{Q}$, we calculate the overall prediction score of the APIs as follows.
\begin{equation}\label{equa:sum_pred}
PredScore_{Q}(i) = \frac{Score_{Q}(i)-Score_{min}}{Score_{max}-Score_{min}} + \mu Relev_{Q}(i)
\end{equation}
where $Score_{Q}(i)$ represents the rank score of the $i$-th API in the recommended list of $Q$, and 
$Relev_{Q}(i)$ is the relevance score of the $i$-th API which takes the position of API into account. $Score_{max}$ and $Score_{min}$ are the maximum and minimum values of the rank score respectively; $\mu$ is the weight which is a dynamic value dependent on the position of the $i$-th API (i.e., $pos_{i}$). In our experiments, $\mu$ is set to $\frac{2}{3\times pos_{i}}$.
We then re-rank $L_Q$ in descending order based on the final prediction score $PredScore_{Q}$. Programmers can choose an adequate API from the re-ranked list corresponding to the query. Meanwhile, the decision will be recorded in the feedback repository.

\section{Evaluation}\label{experiment}

In this section, we evaluate the proposed BRAID approach. We shall mainly study the following research questions (RQs).

\begin{description}
   \item[RQ1] How effective is BRAID to recommend API for  given queries in general? 
   \item[RQ2] How does the feedback information contribute to BRAID for recommending API? In particular, how does the accumulation of the feedback repository improve the performance of BRAID?
%
       \item[RQ3] How do LTR and active learning techniques contribute to BRAID respectively? 
       \item[RQ4] Is the overhead introduced by 
       BRAID acceptable? 
\end{description}



\subsection{Baselines} \label{sec:baseline}
The \BRAID\ approach is essentially an ``add-on" technique, which is designed to be instrumented to \textcolor{black}{existent} query-based API recommendation systems for which we 
use three representative systems, i.e., BIKER, RACK, and NLP2API, as baselines.

BIKER \cite{huang2018api} collects 413 questions, along with their ground-truth APIs, as the testing dataset 
for the empirical study. They are extracted from API-related posts of Stack Overflow following the approach of 
Ye et al.~\cite{ye2016word}. 
The question titles of the posts are considered as the query whereas the APIs referred to in the accepted answers are treated as standard answers. Sometimes, for a common programming task query, \textcolor{black}{if the APIs from other answers which are not marked as accepted ones  are also helpful to solve the problem, human experts are involved to determine whether these APIs should be added to the ground-truth dataset.}

RACK \cite{Rahman2016RACK} collects 150 queries for the evaluation from three Java tutorial sites: KodeJava\footnote{https://kodejava.org}, JavaDB\footnote{https://www.javadb.com} and Java2s\footnote{https://java2s.com}. These sites contain a mass of programming tasks whose descriptions generally are composed of three parts, i.e., a question title, a solution consisting of code snippets, and a comment used to interpret code. Similar to the accepted answers in Stack Overflow posts, the comment explaining the code also refers to one or more APIs which are vital to deal with the question. Hence the ground-truth dataset is made by question titles of the programming tasks in these sites and the corresponding APIs extracted from code interpretation.

NLP2API \cite{rahman2018effective} collects 310 code search query-API pairs. Similar to RACK, the source of data is also the Java tutorial sites. In addition to the sites which RACK refers to, they also focus on the data on CodeJava.\footnote{https://www.codejava.net} Thus, besides the 150 queries already gained by RACK, there are 160 new ground-truth pairs, which make up 310 pairs of NLP2API. Though some query-API pairs of NLP2API are the same as RACK, it has no effect on our evaluation results, \textcolor{black}{since the comparative experiment with each baseline is conducted independently. The query-API pairs of the three baseline work are not merged together, and thus no duplicates would be introduced.} The ground-truth data set of this API recommendation system is composed in the same way as RACK.


%
In the experiments, we reuse the existing datasets, as well as the implementations, from the replication packages of the baselines, i.e., BIKER\footnote{https://github.com/tkdsheep/BIKER-ASE2018}, RACK\footnote{\url{http://homepage.usask.ca/~masud.rahman/rack/}}, and NLP2API\footnote{https://github.com/masud-technope/NLP2API-Replication-Package}. 
\textcolor{black}{In general, to evaluate the performance of machine learning techniques, we follow the standard 10-fold cross validation and  repeat the experiments 5 times. The results are recorded and the average values are calculated as the final results. To avoid bias, the query-API pairs in the feedback repository whose first component is the duplicate of the testing query are removed.} Our implementation is based on XGBoost (ver. 0.82) \cite{chen2016xgboost} and modAL (ver. 0.3.4)\footnote{https://github.com/modAL-python/modAL} for LTR and active learning respectively. \textcolor{black}{XGBoost is an optimized distributed gradient boosting library, implementing machine learning algorithms in the Gradient Boosting framework which can be used, among others, for LTR tasks; ModAL is a modular active learning framework for Python3.} The experiments are conducted on a PC running Windows 10 OS with an AMD Ryzen 5 1600 CPU (6 cores) of 3.2GHz  and 8GB DDR4 RAM.

\textcolor{black}{For the active learning component, oracle has to be utilized. To ensure a fair comparison, we reuse the Stack Overflow posts provided by the baseline tools to build up our oracle. For BIKER, the oracle is based on the 125,847 Stack Overflow posts provided by BIKER after pre-processing. For RACK and NLP2API, note that the dataset of RACK is actually reused by NLP2API, so they share the same oracle, based on the 646,242 Stack Overflow posts provided by NLP2API after pre-processing. 
	In more details, the oracle is in the form of pairs of posts as well as the accepted answers. We extract APIs from the answers by parsing the $<$code$>$ tag. Namely, for each $<$code$>$ tag, we use JDT\footnote{https://www.eclipse.org/jdt/} to construct ASTs based on which  the APIs can be extracted. (We only consider those snippets which can be successfully parsed.) It is common that multiple APIs are present in the code corresponding to one query. We construct a list to include them, which forms the second component  of the pair (i.e., the answer). After this step, we collect 22,041 query-API pairs for BIKER, and 45,943 for RACK and NLP2API. Among these, we further select the pairs based on the following criteria: (1) the question score is positive; (2) the view count exceeds 100. In the end, we obtain 2,434 pairs for BIKER and 3,703 for RACK and NLP2API. To assure the quality of these pairs, we have asked  three researchers in software engineering (including the second author), who are familiar with the context of work, to manually examine 
	the dataset independently and remove those questions not searching for APIs. In case of disagreement, after discussion, the majority-vote strategy is applied to resolve the conflicts. 
} 

\subsection{Performance metrics}
We leverage three widely used metrics in literature (e.g., \cite{silva2019recommending, ye2014, liu2018effective, mcmillan2011portfolio, manning2010introduction}) to measure the performance of our approach.

\begin{itemize}
\item Hit@k/Top-k Accuracy, which is the percentage of queries of which at least one recommended API is relevant within the top $k$ results. Formally,
  \[
    Hit@k=\frac{rel(k)}{|Q|}
  \]
  where $rel(k)$ represents the number of queries whose relevant API appears in the top-$k$, 
  and $|Q|$ is the total number of the queries.

\item Mean Average Precision (MAP) is the mean of the average precision (AP) scores for each query. Formally, 
  \[
    MAP=\frac{1}{|Q|}\sum_{i=1}^{|Q|} AP(i),  AP=\frac{1}{|K|}\sum_{k \in K} \frac{num(k)}{k}
  \]
where $K$ is the set of ranking position of the relevant APIs of the ranked APIs list of the $i$-th query, and $num(k)$ represents the number of relevant API in the top-$k$.

\item Mean Reciprocal Rank (MRR) calculates the inverse of the first appearing relevant API of a query, then adds them up and averages as the result.
 \[ MRR=\frac{1}{|Q|}\sum_{i=1}^{|Q|}\frac{1}{rank_i}\]
 where $rank_i$ represents the ranking position of the first relevant API in the $i$-th query.
\end{itemize}

\textcolor{black}{
\subsection{Statistical tests}
To assess the significance of experiment results, we carry out a statistical analysis on the obtained results. Following the guidelines in~\cite{DBLP:conf/icse/ArcuriB11}, we conduct
the Mann-Whitney U test 
to determine whether the improvement is significant in a statistical sense. Moreover, 
we assess the magnitude of the improvement for which we analyze the  effect size via  Vargha and Delaney's $\hat{A}_{12}$ measure, a standardized non-parametric effect size measure. In general, for two algorithms $A$ and $B$, if $\hat{A}_{12}$ is 0.5, the two algorithms are considered equivalent. If $\hat{A}_{12}$ is greater than 0.5, 
the algorithm $A$ has a higher chance to perform better than the algorithm $B$. $\hat{A}_{12}$ is computed by the following statistics~\cite{EffectSize}:
\begin{equation}\label{effectSize}
\hat{A}_{12}=(R_1/m-(m+1)/2)/n,
\end{equation}
where $R_1$ is the rank sum of the first data group, 
$m$ (resp. $n$) is the number of observations in the first (resp. second) data sample. In our experiments, we run two algorithms the same number of times, i.e., the values of $m$ and $n$ are both set to 5.}

\subsection{Experimental results}
\noindent \textbf{RQ1.
How effective is BRAID to recommend API for given queries in general?}


\textcolor{black}{In the experiment, we randomly 
select 10 query-answer pairs from the training set to build the feedback repository which is fixed for each run of the experiment.} One such example is
given in Table~\ref{tab2}. 
\begin{table}[!h]
	\centering
	\caption{10 queries in feedback repository \label{tab2}}
	\begin{tabular}{|c|}
		\hline
		Query \\
		\hline
		Convert Point coordinates to Screen coordinates in JavaFX?\\
		\hline
		Get the last three chars from any string - Java\\
		\hline
		How to handle if a sql query finds nothing? Using resultset in java\\
		\hline
		Java: Make one item of a jcombobox unselectable\\
		(like for a sub-caption) and edit font of that item\\
		\hline
		Java String to byte conversion is different\\
		\hline
		LinkedBlockingQueue - java - queue full\\
		\hline
		Set JLabel Visible when JButton is clicked in actionPerformed\\
		\hline
		Adding JPanels to regions other than CENTER\\
		\hline
		Sorting based on value of object\\
		\hline
		Simple calculate using inheritance and\\
		Scanner how i handle these Exceptions?\\
		\hline
	\end{tabular}
\end{table}
%
Note that the feedback repository is randomly selected and removed from the testing set.
We fix the feedback repository because the main aim of this experiment is to investigate the effectiveness of the feedback repository to recommendation improvements. 

We use queries from the testing set to evaluate
three baselines BIKER, RACK and NLP2API augmented with BRAID respectively, i.e., BRAID (BIKER), BRAID (RACK) and BRAID (NLP2API).  
The performance is measured by Hit@1, Hit@3, Hit@5, MAP and MRR. For the comparison with BIKER, the recommendation is at both the method level and the class level whereas for RACK and NLP2API, it is at the class level because these two tools are designed to recommend API classes only.



For each experiment, we carry out 10-fold cross validation. Namely, we randomly split the dataset by 9:1, and each time one fold is used as the testing data while the remaining  nine folds are used to train the LTR model. We calculate the average metrics of 10 times. Such an experiment is repeated 5 times. For each run, the feedback repository is again updated with 10 randomly selected pairs among the remaining ones. Then the average of the five experiments is taken as the final result shown in Table~\ref{tab3}.


\begin{table}[!h]
	\centering
	\caption{Evaluation results comparison (BRAID vs. baselines) with fixed feedback repository ('Abs. imp.' stands for 'absolute improvement'; 'rel. imp.' stands for 'relative improvement') \label{tab3}}
	\setlength{\tabcolsep}{0.7mm}{
		\begin{tabular}{|c|c|c|c|c|c|c|}
			\hline
			Baseline & Technique & Hit@1 & Hit@3 & Hit@5 & MAP & MRR \\
			\hline
			\multirow{3}{1.2cm}{BIKER (Method Level)}  
			& Original & 0.4231 & 0.6607 & 0.7747 & 0.5534 & 0.5685 \\
			& Avg. BRAID & \textbf{0.4401} & \textbf{0.6718} & \textbf{0.7800} & \textbf{0.5652} & \textbf{0.5808} \\
			& Abs. Imp. & 1.70\% & 1.11\% & 0.52\% & 1.18\% & 1.23\% \\
			& Rel. Imp. & 4.02\% & 1.68\% & 0.68\% & 2.14\% & 2.16\% \\
			\hline
			\multirow{3}{1.2cm}{BIKER (Class Level)}  
			& Original & 0.5472 & 0.8136 & 0.9031 & 0.6753 & 0.6522 \\
			& Avg. BRAID & \textbf{0.5616} & \textbf{0.8173} & \textbf{0.9046} & \textbf{0.6890} & \textbf{0.7031} \\
			& Abs. Imp. & 1.44\% & 0.38\% & 0.14\% & 1.37\% & 5.09\% \\
			& Rel. Imp. & 2.62\% & 0.46\% & 0.16\% & 2.03\% & 7.81\% \\
			\hline
			\multirow{3}{1.2cm}{RACK}  
			& Original & 0.3267 & 0.5133 & 0.6267 & 0.4203 & 0.4506 \\
			& Avg. BRAID & \textbf{0.3853} & \textbf{0.5493} & \textbf{0.6453} & \textbf{0.4630} & \textbf{0.4935} \\
			& Abs. Imp. & 5.87\% & 3.60\% & 1.87\% & 4.28\% & 4.29\% \\
			& Rel. Imp. & 17.96\% & 7.01\% & 2.98\% & 10.17\% & 9.52\% \\
			\hline
			\multirow{3}{1.2cm}{NLP2API}  
			& Original & 0.3516 & 0.5323 & 0.6000 & 0.4111 & 0.4604 \\
			& Avg. BRAID & \textbf{0.3716} & \textbf{0.5477} & \textbf{0.6071} & \textbf{0.4336} & \textbf{0.4740} \\
			& Abs. Imp. & 2.00\% & 1.55\% & 0.71\% & 2.25\% & 1.36\% \\
			& Rel. Imp. & 5.69\% & 2.91\% & 1.18\% & 5.48\% & 2.95\% \\
			\hline
	\end{tabular}}
\end{table}

From Table~\ref{tab3}, one can see that  almost all metrics have improved compared with the baselines. In general, even when a small-scale  feedback repository (with merely 10 pairs) is harnessed, 
BRAID demonstrates 
the relative
improvements over the  baselines by 4.02\%, 1.68\%, 0.68\%, 2.14\%, 2.16\% for BIKER (method level), 2.62\%, 0.46\%, 0.16\%, 2.03\%, 7.81\% for BIKER (class level), 17.96\%, 7.01\%, 2.98\%, 10.17\%, 9.52\% for RACK and 5.69\%, 2.91\%, 1.18\%, 5.48\%, 2.95\% for NLP2API respectively. In addition, a statistical analysis of the results have been carried out. 
We applied the Mann-Whitney U test to the results. BRAID and each of the three baselines are considered as pair groups. For each run, we collect the average values of the above metrics as outcomes. The experiment is repeated 5 times, and we obtain a sample size of 5 for each pair group. Since multiple comparisons are conducted in terms of Hit@1, Hit@3, Hit@5, MAP, and MRR, we adopt the Bonferroni correction. 
In a nutshell, if the significance level is set to be $\alpha$, and $m$ individual tests are performed, the null hypothesis can be rejected only if the $p-$value is less than the adjusted threshold $\alpha/m$. In our experiment, we follow the convention that $\alpha=0.05$. The number of comparisons is 5, hence the adjusted threshold is 0.01.
For the Mann-Whitney U test, 
the $p-$values are all less than $0.005$, which indicates that the improvements are statistically significant at the confidence level of $95\%$. The Vargha and Delaney $\hat{A}_{12}$ is 1, which represents the highest effect size. 
This confirms that the feedback repository is effective in boosting the performance of API recommendations.
In addition, the same feedback repository works well on the three API recommendation systems (BIKER, RACK and NLP2API), which demonstrates the  generalization ability of BRAID for query-based API recommendation.

\medskip
\noindent\textbf{RQ2. How does the accumulation of the feedback repository improve the performance of BRAID?}
%
\begin{table*}
	\centering
	\caption{Evaluation results comparison with accumulated feedback repository\label{tab_braid}}
	\setlength{\tabcolsep}{2.85mm}{
		\begin{tabular}{|c|c|c|c|c|c|c|c|c|c|c|c|c|}
			\hline
			Baseline & Metric & Original & 10\% & 20\% & 30\% & 40\% & 50\% & 60\% & 70\% & 80\% & 90\% & 100\% \\
			\hline
			\multirow{5}{1.2cm}{BIKER (Method Level)}
			& Hit@1 & 0.4231 & 0.4418 & 0.4704 & 0.4931 & 0.4986 & 0.5020 & 0.5073 & 0.5112 & 0.5146 & 0.5170 & 0.5175 \\
			& Hit@3 & 0.6607 & 0.6815 & 0.7018 & 0.7140 & 0.7178 & 0.7178 & 0.7193 & 0.7203 & 0.7203 & 0.7208 & 0.7213 \\
			& Hit@5 & 0.7747 & 0.7825 & 0.7945 & 0.8024 & 0.8062 & 0.8067 & 0.8072 & 0.8077 & 0.8091 & 0.8096 & 0.8110 \\
			& MAP   & 0.5534 & 0.5689 & 0.5919 & 0.6072 & 0.6106 & 0.6128 & 0.6155 & 0.6176 & 0.6205 & 0.6214 & 0.6223 \\
			& MRR   & 0.5685 & 0.5816 & 0.6035 & 0.6189 & 0.6226 & 0.6252 & 0.6282 & 0.6308 & 0.6334 & 0.6346 & 0.6356 \\
			\hline
			\multirow{5}{1.2cm}{BIKER (Class Level)}
			& Hit@1 & 0.5472 & 0.5647 & 0.5749 & 0.5857 & 0.6011 & 0.6016 & 0.6021 & 0.6083 & 0.6102 & 0.6132 & 0.6151 \\
			& Hit@3 & 0.8136 & 0.8185 & 0.8195 & 0.8316 & 0.8317 & 0.8330 & 0.8355 & 0.8417 & 0.8433 & 0.8447 & 0.8500 \\
			& Hit@5 & 0.9031 & 0.9004 & 0.9052 & 0.9058 & 0.9063 & 0.9070 & 0.9072 & 0.9072 & 0.9077 & 0.9082 & 0.9107 \\
			& MAP   & 0.6753 & 0.6878 & 0.6914 & 0.7021 & 0.7128 & 0.7142 & 0.7143 & 0.7186 & 0.7205 & 0.7214 & 0.7245 \\
			& MRR   & 0.6522 & 0.7051 & 0.7099 & 0.7197 & 0.7279 & 0.7285 & 0.7291 & 0.7343 & 0.7352 & 0.7368 & 0.7394 \\
			\hline
			\multirow{5}{1.2cm}{RACK}
			& Hit@1 & 0.3267 & 0.4160 & 0.4587 & 0.4827 & 0.4840 & 0.4893 & 0.4907 & 0.4947 & 0.5000 & 0.5040 & 0.5067 \\
			& Hit@3 & 0.5133 & 0.5680 & 0.5933 & 0.6013 & 0.6120 & 0.6133 & 0.6147 & 0.6173 & 0.6200 & 0.6360 & 0.6400 \\
			& Hit@5 & 0.6267 & 0.6453 & 0.6640 & 0.6667 & 0.6720 & 0.6733 & 0.6733 & 0.6773 & 0.6813 & 0.6813 & 0.6867 \\
			& MAP   & 0.4203 & 0.4789 & 0.5211 & 0.5345 & 0.5418 & 0.5434 & 0.5434 & 0.5490 & 0.5538 & 0.5588 & 0.5620 \\
			& MRR   & 0.4506 & 0.5120 & 0.5455 & 0.5622 & 0.5654 & 0.5675 & 0.5692 & 0.5722 & 0.5765 & 0.5819 & 0.5852 \\
			\hline
			\multirow{5}{1.2cm}{NLP2API}
			& Hit@1 & 0.3516 & 0.3871 & 0.4452 & 0.4761 & 0.4916 & 0.5181 & 0.5213 & 0.5226 & 0.5258 & 0.5310 & 0.5355 \\
			& Hit@3 & 0.5323 & 0.5561 & 0.5877 & 0.6039 & 0.6187 & 0.6284 & 0.6316 & 0.6342 & 0.6348 & 0.6348 & 0.6355 \\
			& Hit@5 & 0.6000 & 0.6187 & 0.6413 & 0.6426 & 0.6523 & 0.6555 & 0.6619 & 0.6626 & 0.6632 & 0.6645 & 0.6645 \\
			& MAP   & 0.4111 & 0.4451 & 0.4851 & 0.5123 & 0.5249 & 0.5408 & 0.5450 & 0.5480 & 0.5482 & 0.5524 & 0.5549 \\
			& MRR   & 0.4604 & 0.4885 & 0.5290 & 0.5502 & 0.5627 & 0.5807 & 0.5841 & 0.5867 & 0.5881 & 0.5912 & 0.5937 \\
			\hline
	\end{tabular}}
\end{table*}

In the first experiment, we fix the feedback repository. In real scenarios, the feedback repository is to be updated with the feedback received from the end users. 
How does the accumulation of the feedback repository (representing the feedback information) influence the recommendation results?
Our experiment aims to answer this question.

We randomly select the query-answer pairs
from the training set 
to form the feedback repository. The size of the feedback repository varies from 0\% to 100\% of the training set, with an increment of 10\%. 
Note that the baseline is represented by the case of size equal to 0\%, where the feedback repository is disabled. 
\textcolor{black}{For each sampled feedback repository, as before, we carry out 10-fold cross validation which is repeated 5 times and the reported results represent the average.}


\begin{figure*}
    \subfigure[The performance of BIKER (Method Level)]{
	\includegraphics[width=0.23\hsize]{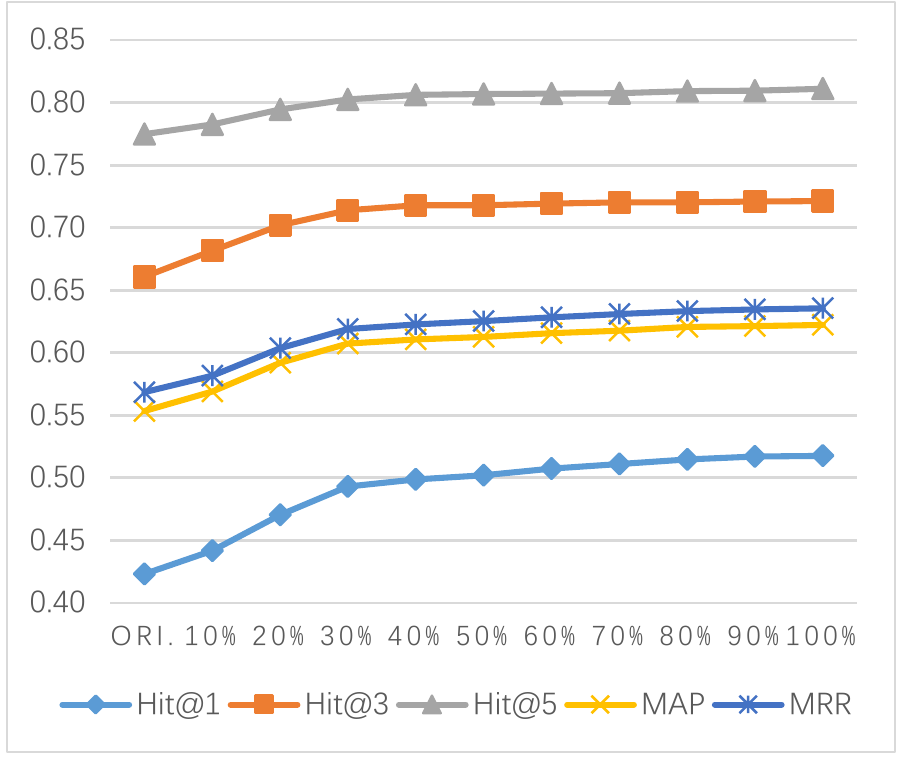}}
    \hspace{1ex}
    \subfigure[The performance of BIKER (Class Level)]{
    \includegraphics[width=0.23\hsize]{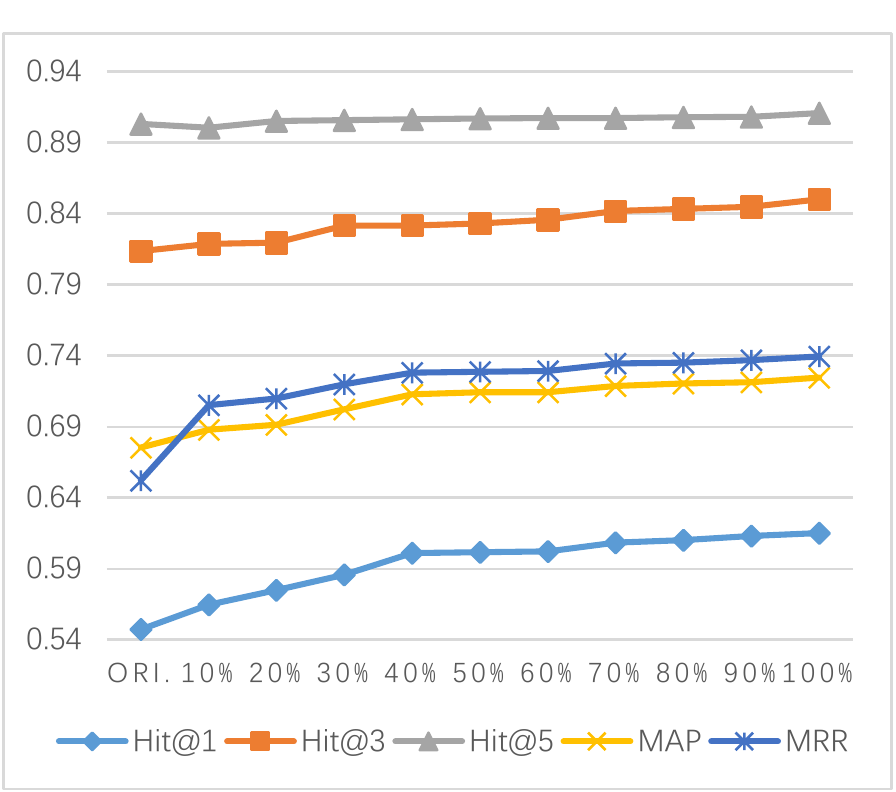}}
    \hspace{1ex}
    \subfigure[The performance of RACK]{
	\includegraphics[width=0.23\hsize]{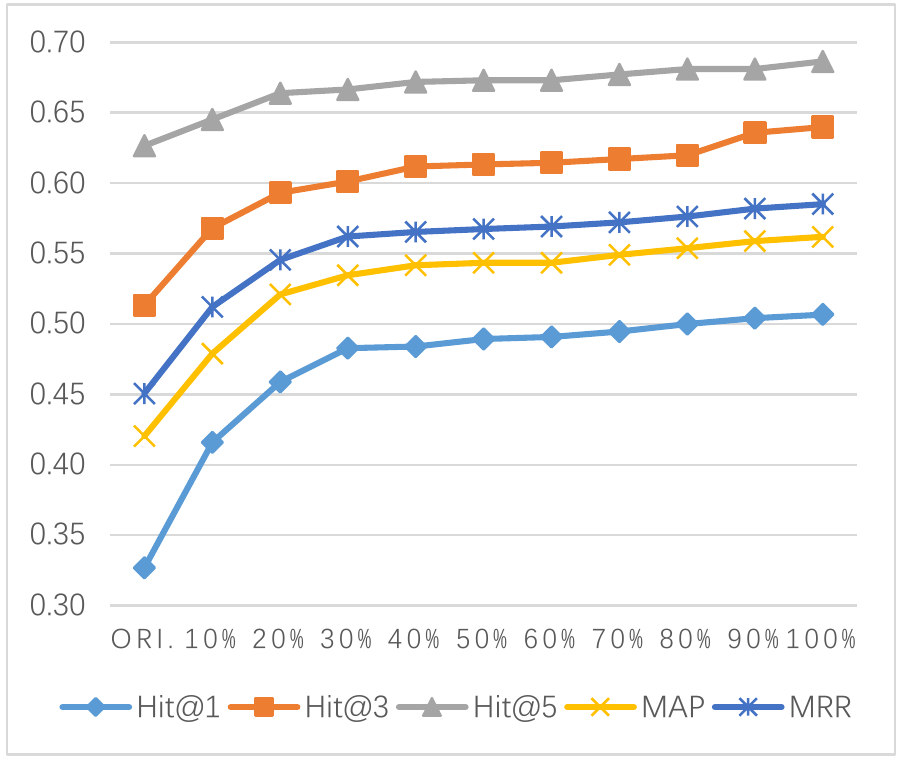}}
    \hspace{1ex}
    \subfigure[The performance of NLP2API]{
	\includegraphics[width=0.23\hsize]{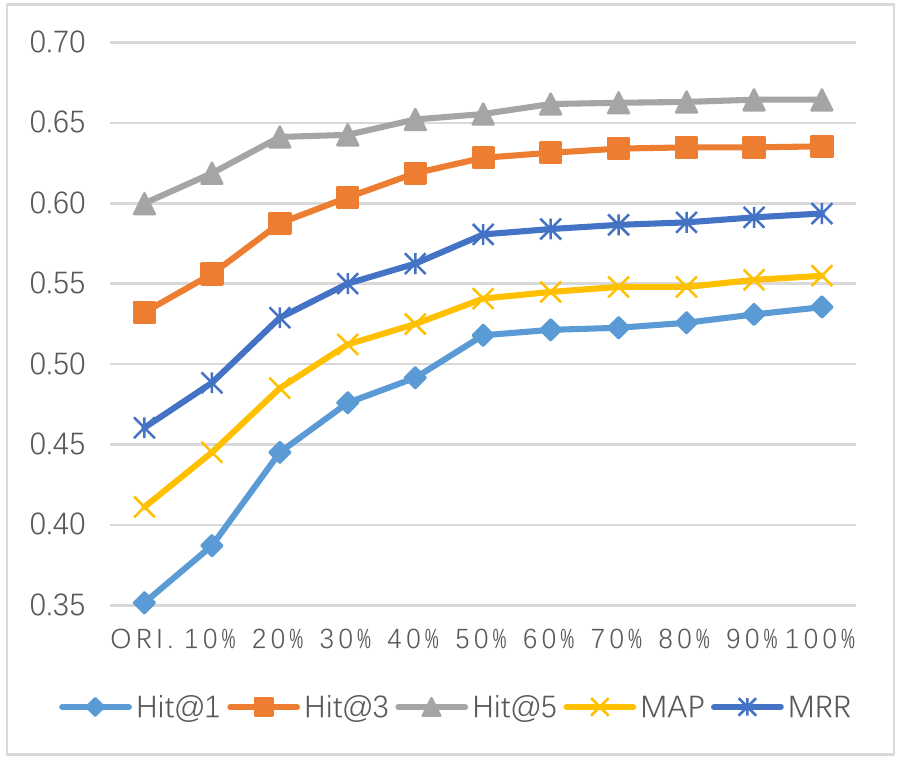}}
\caption{Learning curves of BRAID with feedback information for baselines \label{fig:perform}}
\end{figure*}


Table~\ref{tab_braid} presents the experimental results. To better visualize the trend, we also plot the results in Fig.~\ref{fig:perform}.
One can observe that the performance improves with the accumulation of the feedback repository. This is consistent across all the three baselines, indicating the generalizability of our approach for query-based recommendation.
%
In particular, all the metrics have been enhanced considerably. 
\textcolor{black}{
The MAP and MRR are 6\% up for BIKER at the method level, over 5\% up for BIKER at the class level, over 13\% up for RACK and NLP2API.}

\begin{figure*}
    \subfigure[BIKER Method Level Hit@1]{
	\includegraphics[width=0.23\hsize]{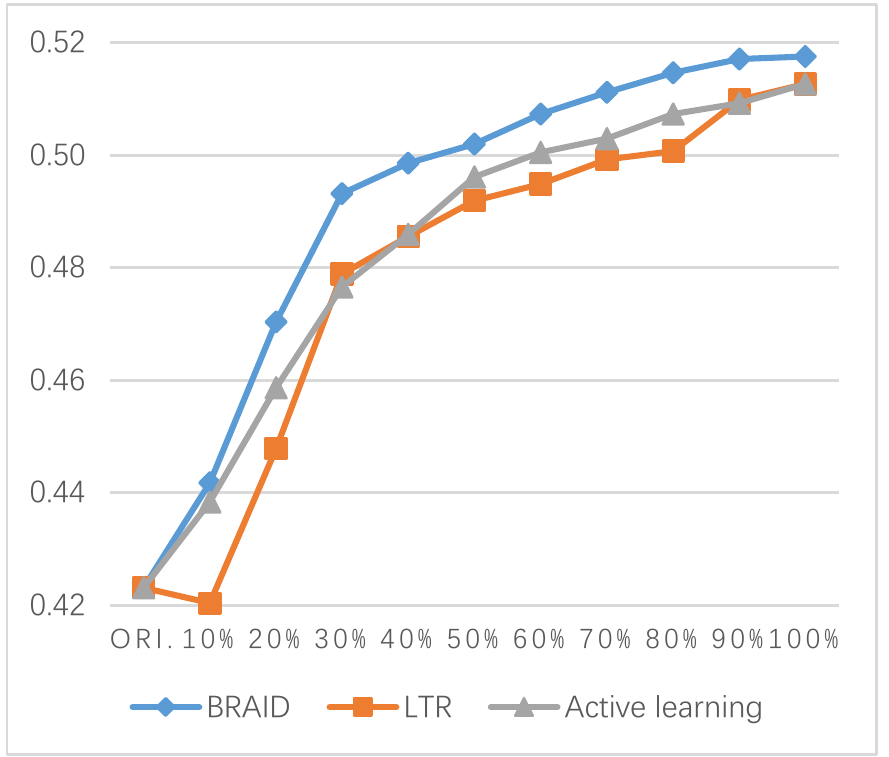}}
	\subfigure[BIKER Class Level Hit@1]{
	\includegraphics[width=0.23\hsize]{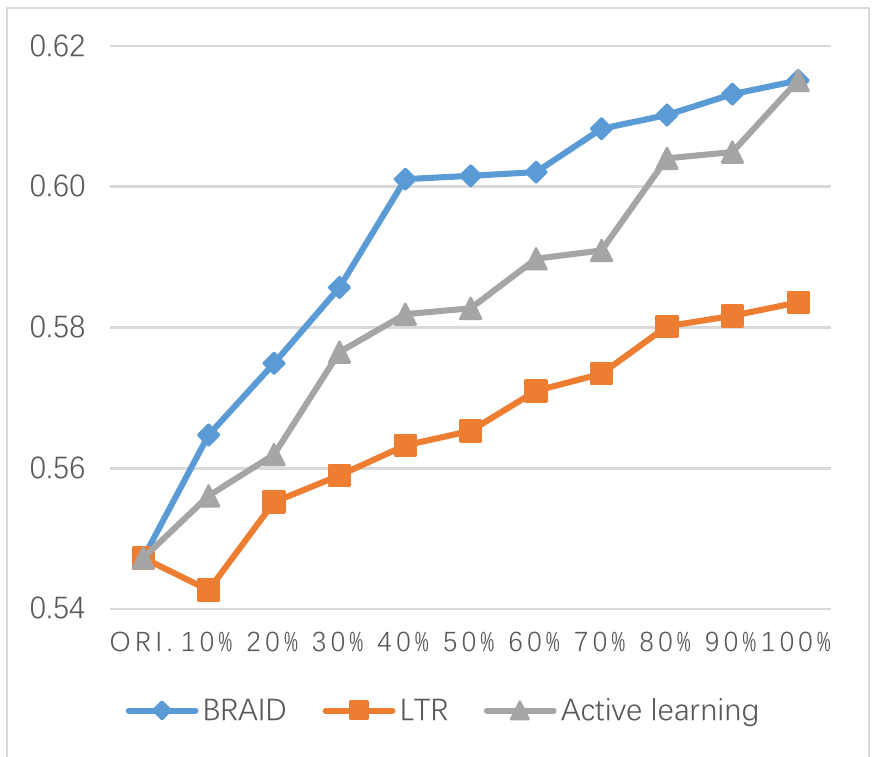}}
    \hspace{1ex}
    \subfigure[RACK Hit@1]{
	\includegraphics[width=0.23\hsize]{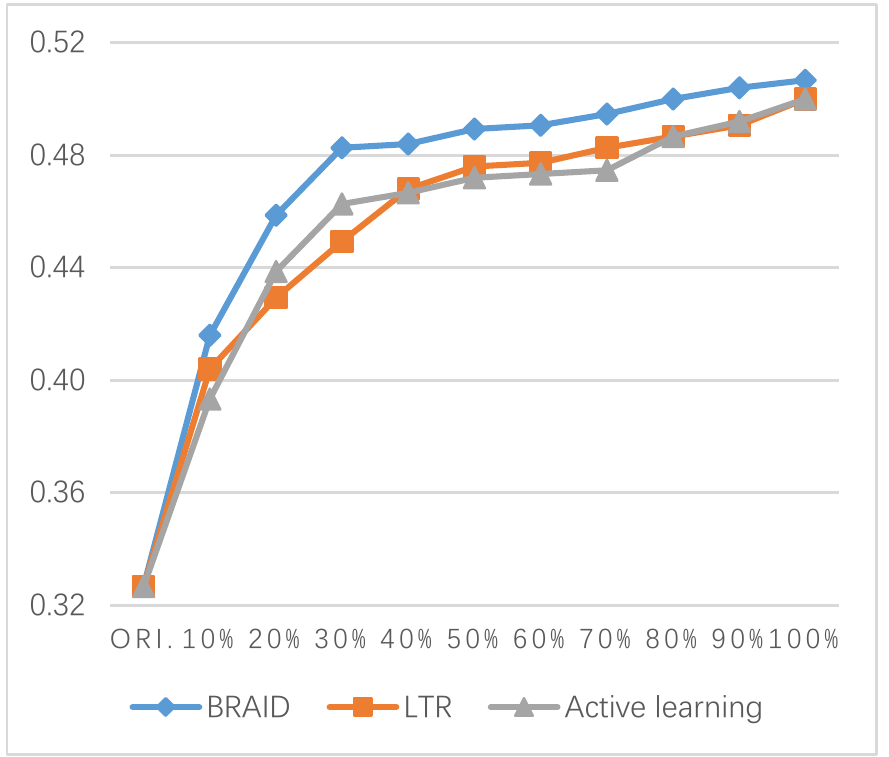}}
    \hspace{1ex}
    \subfigure[NLP2API Hit@1]{
	\includegraphics[width=0.23\hsize]{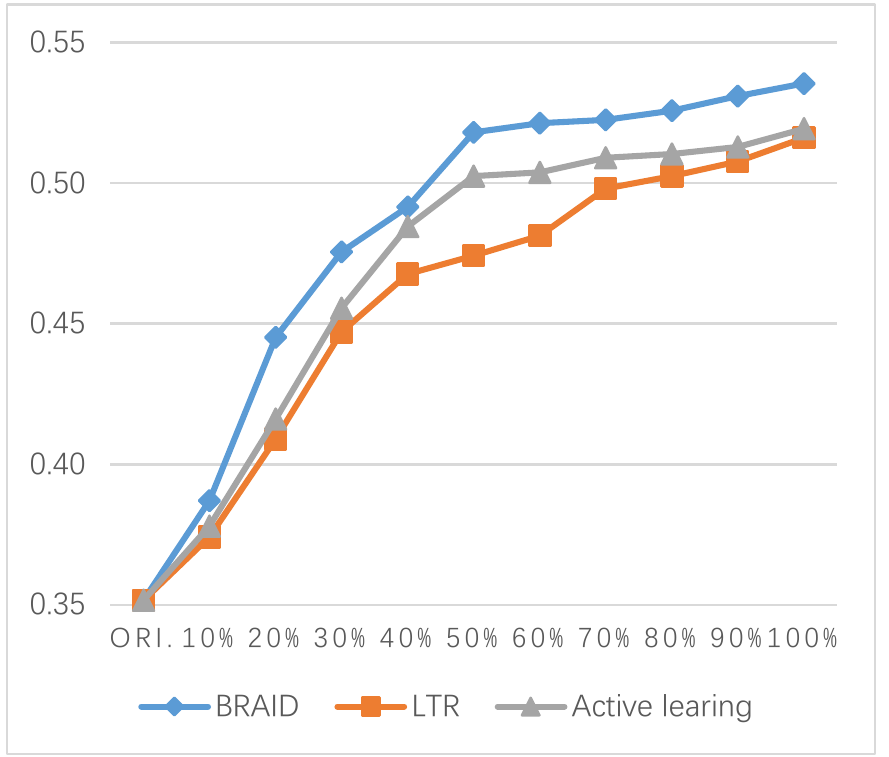}\label{fig:sub}}
\caption{The performance metrics of Baselines Hit@1 \label{fig:hit1}}
\end{figure*}

Arguably, the most important  indicator Hit@1 enjoys the largest boosting, which demonstrates that our approach can rank the most relevant API to the top-1 through feedback information. Fig.~\ref{fig:hit1} 
shows the Hit@1 metric of all three baselines: \textcolor{black}{Hit@1 is increased by 9.44\% for BIKER (method level), by 6.79\% for BIKER (class level), by 18\% for RACK, and by 18.39\% for NLP2API. Moreover, we use the Mann-Whitney U test and  Vargha and Delaney's  $\hat{A}_{12}$ statistic to examine these experimental results. Most $p-$values are in the range of $0.003$ to $0.005$, with effect size $1$, indicating that the improvements are statistically significant at the confidence level of $95\%$. However, for BIKER (method level) there were 2 cases (metrics Hit@3 and hit@5 for 10\% size of feedback repository) out of 50 where the $p-$values were higher than 0.01 (i.e., the adjusted threshold with the Bonferroni correction). 
For BIKER (class level) there were 3 cases (metrics hit@5 for 10\%, 20\% and 30\% size of feedback repository) out of 50 where the $p-$values were higher than 0.01 (i.e., the adjusted threshold with the Bonferroni correction). 
For NLP2API, there was also one case (i.e., metrics Hit@5 for 10\% size of feedback repository) where the $p$-value is higher than 0.01.   
We suspect that, when the feedback information is insufficient, our approach may not bring significant improvement on certain occasions.  However, with the growth of feedback, our approach does show significant improvement over the baselines.}

To further demonstrate how the user is involved and the effectiveness of our approach, we conduct a further experiment where we consider a pseudo-user. 
We randomly select 50 queries, and the pseudo-user is programming during which the 50 queries are to be made. 
During each query, BRAID 
recommends APIs based on the feedback repository, and the pseudo-user selects API(s).
The query and selected API(s) are used to expand the feedback repository. We train the models as soon as the feedback repository is not empty. The model is not re-trained during the 50 queries. 
Table~\ref{tab：imitate} shows the results for pseudo-user experiment. The conclusion is consistent with other experiments that the results of Hit@1 metric improve the most. For example, Hit@1 increase for NLP2API is around 5\%, and for RACK is over 9\%.
\begin{table}[h]
	\centering
	\caption{Evaluation results comparison with a pseudo-user\label{tab：imitate}}
	\setlength{\tabcolsep}{0.7mm}{
		\begin{tabular}{|c|c|c|c|c|c|c|}
			\hline
			Baseline & Technique & Hit@1 & Hit@3 & Hit@5 & MAP & MRR \\
			\hline
			\multirow{3}{1.2cm}{BIKER (Method Level)}  
			& Original & 0.4213 & 0.6543 & 0.7639 & 0.5412 & 0.5496 \\
			& Avg. BRAID & \textbf{0.4800} & \textbf{0.7000} & \textbf{0.8000} & \textbf{0.5924} & \textbf{0.5967} \\
			& Abs. Imp. & 5.87\% & 4.57\% & 3.61\% & 5.12\% & 4.71\% \\
			& Rel. Imp. & 13.94\% & 6.98\% & 4.73\% & 9.46\% & 8.56\% \\
			\hline
			\multirow{3}{1.2cm}{BIKER (Class Level)}  
			& Original & 0.5373 & 0.8064 & 0.8961 & 0.6713 & 0.6851 \\
			& Avg. BRAID & \textbf{0.5600} & \textbf{0.8200} & \textbf{0.9000} & \textbf{0.6783} & \textbf{0.7054} \\
			& Abs. Imp. & 2.27\% & 1.36\% & 0.39\% & 0.70\% & 2.03\% \\
			& Rel. Imp. & 4.22\% & 1.69\% & 0.44\% & 1.04\% & 2.96\% \\
			\hline
			\multirow{3}{1.2cm}{RACK}  
			& Original & 0.3233  & 0.5067  & 0.6067  & 0.4150  & 0.4421  \\
			& Avg. BRAID & \textbf{0.4200 } & \textbf{0.6000 } & \textbf{0.6600 } & \textbf{0.5155 } & \textbf{0.5410 } \\
			& Abs. Imp. & 9.67\% & 9.33\% & 5.33\% & 10.05\% & 9.89\% \\
			& Rel. Imp. & 29.91\% & 18.41\% & 8.79\% & 24.22\% & 22.38\% \\			
			\hline
			\multirow{3}{1.2cm}{NLP2API}  
			& Original & 0.3528 & 0.5355 & 0.6065 & 0.4155 & 0.4627 \\
			& Avg. BRAID & \textbf{0.4000} & \textbf{0.5600} & \textbf{0.6400} & \textbf{0.4643} & \textbf{0.5072} \\
			& Abs. Imp. & 4.72\% & 2.45\% & 3.35\% & 4.88\% & 4.45\% \\
			& Rel. Imp. & 13.38\% & 4.58\% & 5.52\% & 11.73\% & 9.62\% \\
			\hline
	\end{tabular}}
\end{table}


\medskip
\noindent\textbf{RQ3. How do LTR and active learning techniques contribute to BRAID respectively?}

Recall that our approach makes use of two learning techniques, i.e., LTR and active learning. To better interpret the performance improvement of \BRAID, we perform an ablation analysis to pinpoint the individual contribution of each technique.

In the experiment, similar to the previous one, we gradually increase the size of the feedback repository. At each stage, we disable either LTR or active learning and collect the performance metrics accordingly.
We calculate the results of baselines for testing data and  the averages (over all stages) of LTR and active learning techniques respectively. 
The experimental results are given in Table \ref{tab:aver}.

\begin{table}
	\centering
	\caption{Evaluation results for our framework comparing with baselines ('AL' stands for active learning) \label{tab:aver}}
	\setlength{\tabcolsep}{0.6mm}{
		\begin{tabular}{|c|c|c|c|c|c|c|}
			\hline
			Approach & Technique & Hit@1 & Hit@3 & Hit@5 & MAP & MRR \\
			\hline
			\multirow{7}{1.2cm}{BIKER (Method Level)}  
			& Original & 0.4231 & 0.6607 & 0.7747 & 0.5534 & 0.5685 \\
			& Avg. LTR & 0.4842 & 0.7047 & 0.8002 & 0.6002 & 0.6116 \\
			& Avg. AL & 0.4888 & 0.7044 & 0.7959 & 0.6013 & 0.6151 \\
			& Avg. BRAID & \textbf{0.4974} & \textbf{0.7135} & \textbf{0.8037} & \textbf{0.6089} & \textbf{0.6214} \\
			& Rel. Imp. LTR & 14.43\% & 6.65\% & 3.28\% & 8.46\% & 7.58\% \\
			& Rel. Imp. AL & 15.53\% & 6.61\% & 2.73\% & 8.65\% & 8.20\% \\
			& Rel. Imp. BRAID & 17.55\% & 7.98\% & 3.74\% & 10.02\% & 9.31\% \\
			\hline
			\multirow{7}{1.2cm}{BIKER (Class Level)}  
			& Original & 0.5472 & 0.8136 & 0.9031 & 0.6753 & 0.6522 \\
			& Avg. LTR & 0.5675 & 0.8047 & 0.8937 & 0.6848 & 0.7010 \\
			& Avg. AL & 0.5864 & 0.8321 & 0.9041 & 0.7044 & 0.7193 \\
			& Avg. BRAID & \textbf{0.5977} & \textbf{0.8349} & \textbf{0.9066} & \textbf{0.7108} & \textbf{0.7266} \\
			& Rel. Imp. LTR & 3.71\% & -1.09\% & -1.05\% & 1.41\% & 7.49\% \\
			& Rel. Imp. AL & 7.16\% & 2.28\% & 0.11\% & 4.31\% & 10.30\% \\
			& Rel. Imp. BRAID & 9.22\% & 2.63\% & 0.38\% & 5.25\% & 11.41\% \\
			\hline
			\multirow{7}{1.2cm}{RACK}  
			& Original & 0.3267 & 0.5133 & 0.6267 & 0.4203 & 0.4506 \\
			& Avg. LTR & 0.4664 & 0.6060 & 0.6701 & 0.5254 & 0.5529 \\
			& Avg. AL & 0.4660 & 0.5828 & 0.6597 & 0.5249 & 0.5485 \\
			& Avg. BRAID & \textbf{0.4827} & \textbf{0.6116} & \textbf{0.6721} & \textbf{0.5387} & \textbf{0.5638} \\
			& Rel. Imp. LTR & 42.78\% & 18.05\% & 6.94\% & 25.02\% & 22.68\% \\
			& Rel. Imp. AL & 42.65\% & 13.53\% & 5.28\% & 24.90\% & 21.71\% \\
			& Rel. Imp. BRAID & 47.76\% & 19.14\% & 7.26\% & 28.17\% & 25.11\% \\
			\hline
			\multirow{7}{1.2cm}{NLP2API}  
			& Original & 0.3516 & 0.5323 & 0.6000 & 0.4111 & 0.4604 \\
			& Avg. LTR & 0.4678 & 0.5917 & 0.6386 & 0.5061 & 0.5434 \\
			& Avg. AL & 0.4792 & 0.6064 & 0.6405 & 0.5153 & 0.5532 \\
			& Avg. BRAID & \textbf{0.4954} & \textbf{0.6166} & \textbf{0.6527} & \textbf{0.5257} & \textbf{0.5655} \\
			& Rel. Imp. LTR & 33.05\% & 11.18\% & 6.43\% & 23.11\% & 18.02\% \\
			& Rel. Imp. AL & 36.29\% & 13.93\% & 6.74\% & 25.34\% & 20.15\% \\
			& Rel. Imp. BRAID & 40.90\% & 15.84\% & 8.78\% & 27.87\% & 22.81\% \\
			\hline
	\end{tabular}}
\end{table}

From the table, we can see the roles that learning-to-rank and active learning techniques have played in boosting the API recommendation.
These two techniques make different contributions in all of the baselines, especially at different stages. 
Moreover, 
the performance of RACK is the lowest among the three baselines, but gets the highest boost with our approach.
The improvement tendency of two techniques is consistent for 
all the three baselines. 
We also find, from the improvement  trend of the three baselines, that both techniques focus more on the Hit@1, MAP, MRR and Hit@3 than Hit@5. Among them, the effect of Hit@1 is outstanding.  Despite LTR and active learning techniques optimize the performance in different ways, overall neither of them perform better than the joint force, which justifies the methodology adopted by BRAID.



In Fig.~\ref{fig:hit1}, we plot the Hit@1 curves of the overall BRAID approach (as discussed in \textbf{RQ2}), LTR and active learning with respect to feedback sizes. From the figures, we can see that when the data of feedback repository is small, active learning performs better (except RACK). When there is a lot of feedback data, LTR performs better on RACK and NLP2API. 
With the greater engagement of feedback, in general, LTR, active learning and BRAID all grow steadily and perform better than the original baselines. (The Hit@1 metrics of BIKER (method level), BIKER (class level), RACK, NLP2API are 42.31\%, 54.72\%, 32.67\%, 35.16\% respectively.) It is noteworthy that the overall BRAID achieves the greatest improvement 
which confirms the importance of joint force of LTR and active learning. 

\medskip
\noindent\textbf{RQ4. Is the computational overhead introduced by BRAID acceptable?}

As an ``add-on" technique, when used in conjunction with existing recommendation systems,  \BRAID\ boosts the effectiveness (as demonstrated by the previous experiments) but inevitably introduces overheads. Are these overheads acceptable? This is what we are investigating.


Table~\ref{tab:runtime} shows the runtime of our approach. The original time records the runtime of the baseline. The extraction time represents the time spent on feature extraction. The training time represents the time for training the ranking model of BRAID. The ranking time represents the time to re-rank the API recommended list. The total time is the sum of the extraction, training and ranking time, which represents the overhead introduced by BRAID. The pct.(\%) calculates the percentage of the total time in the original time.

We repeat this experiment  for 5 times on each baseline. For each time, we conduct 10 user queries and calculate the runtime of each query. 
From Table~\ref{tab:runtime}, we can see that  most of the total time is spent on training the ranking model while the re-ranking process is largely negligible (measured in seconds). Among the three baselines, BIKER takes the longest time, (14.29 seconds for the method level, 14.11 seconds for the class level), because loading data takes up most of the time.
%
%

Overall, on average BRAID takes 0.2578 seconds on BIKER (method level) , which is 1.8\% more of the original time, 0.2634 seconds on BIKER (class level) , which is 1.87\% more of the original time, 0.118 seconds on RACK, which is 1.18\% more of the original time,  and 0.1001 seconds on NLP2API, which is 2.52\% more of the original time.

\begin{table}[!h]
    \centering
	\caption{ Runtime overhead results\label{tab:runtime}}
    \setlength{\tabcolsep}{0.2mm}{
	\begin{tabular}{|c|c|c|c|c|c|c|}
		\hline
            \multirow{2}{*}{Approach} & \multirow{2}{*}{Original(s)}&  \multicolumn{5}{c|}{Overheads introduced by BRAID}  \\
            \cline{3-7}
            & & Extraction(s) & Training(s) & Ranking(s) &  Total(s) & Pct.(\%)\\
            \hline
            \tabincell{c}{BIKER\\ (Method\\ Level)} & 14.29 & 0.1876 & 0.0697 & 0.0005 & 0.2578 & 1.80\% \\
            \hline
            \tabincell{c}{BIKER\\ (Class\\ Level)} & 14.11 & 0.1862 & 0.0768 & 0.0004 & 0.2634 & 1.87\% \\
            \hline
            RACK & 10 & 0.0871 & 0.0304 & 0.0004 & 0.1180 & 1.18\% \\
            \hline
            NLP2API & 3.97 & 0.0739 & 0.0259 & 0.0004 & 0.1001 & 2.52\% \\
            \hline
    \end{tabular}}
\end{table}

\section{Threats to Validity}\label{threats}
Threats to internal validity are related to experimental errors and biases \cite{Feldt2010Validity}. 
The main threats of this kind originate from the potential bias introduced in the data. \textcolor{black}{To ensure a fair comparison with the baselines, we use the same data published as the replication packages of the original work.} Moreover, we directly employ their tools to avoid possible errors during re-implementation. \textcolor{black}{The experiments in our study are conducted five times, each of which 10-fold cross validation is performed, and the average values are used as the final results.} In the active learning process, we leverage crowdsourced knowledge from Stack Overflow posts as oracles to provide feedback data. This strategy is adopted in many studies, including the comparative study~\cite{rahman2018effective}, and other research work~\cite{nie16}. To ensure the quality, \textcolor{black}{three software engineering researchers have been recruited to double check the extracted data manually and to confirm the correctness of the  labels.}

Threats to external validity focus on the efficacy that the results can be generalized to other cases different from those used in the experiments \cite{Feldt2010Validity}. Indeed, like other empirical studies, it is hard to guarantee that our framework works well on any other third-party recommendation approach. However, we 
believe that the three state-of-the-art tools selected to demonstrate the advantage of our approach are representative, and the comprehensive experiments can well illustrate the performance enhancement. 
In addition, in our experiments, we concentrate on APIs in Java, which is the same strategy adopted in baseline work. Nevertheless, BRAID is designed to be a language-independent framework where our methodology does not capitalize any peculiarities of Java whereby 
we believe it can be adapted to other programming languages than Java.

\section{Related Work}\label{related}

%

Recommendation systems have been intensively studied in software engineering to assist developers with a wide range of activities~\cite{robillard09,gasparic16}. Rather than a detailed literature review, 
we shall mainly discuss those closely related with ours. Particularly, we focus on three threads of work, i.e., search based code recommendation, generation based code recommendation/completion and results ranking related techniques.

\smallskip
\noindent\textbf{Search based code recommendation.} Code recommendation generally starts from code search. When facing a programming problem, developers usually turn to the Internet for help. Indeed, a recent case study conducted at Google confirmed that developers search for code very frequently~\cite{sadowski15}. Work of this category typically leverages code from open source projects, sometimes augmented with various software artifacts to enhance recommendation precision. Examples include  Strathcona~\cite{holmes05}, Portfolio~\cite{mcmillan2011portfolio}, BCC~\cite{hou2011evaluation}, DroidAssist~\cite{nguyen2015recommending}, SENSORY~\cite{ai2019sensory}, and Aroma~\cite{luan19}. Strathcona recommends code examples for developers by comparing structural similarity in the code repository; Portfolio mainly combines NLP, PageRank~\cite{Brin1998The} and spreading activation network algorithms to find the most relevant code for users; BCC leverages a set of strategies to suggest API candidates, including type-based sorting, filtering, and grouping; DroidAssist uses code context including the current method calls to infer and recommend the following APIs; SENSORY considers the statement sequence information and uses the Burrows-Wheeler Transform algorithm to search in the code repository, and then re-rank the result based on the structure information; Aroma takes a partial code snippet as query input, and returns a set of code snippets as recommendations.  The above approaches mainly rely on code information to perform recommendation. 

Meanwhile, some approaches employ additional information from other software artifacts or crowdsourced knowledge. Examples include BIKER~\cite{huang2018api}, RACK~\cite{Rahman2016RACK}, and NLP2API~\cite{rahman2018effective}, all of which serve as our baselines in this paper. These approaches leverage Q\&A posts from Stack Overflow website to find the most relevant APIs. 
NLP2API also incorporates (pseudo-) feedback information as our work, but its purpose is to reformulate the query. Similarly, QUICKAR~\cite{rahman2016quickar} also aims to automatically provide reformulation of a given query. Some examples augmented with other information for recommendation are APIREC~\cite{nguyen2016api}, and FOCUS~\cite{nguyen2019focus}. APIREC leverages fine-grained change commit history from Github to extract frequent change patterns to supplement the recommendation process. FOCUS tackles the usage pattern recommendation problem from the perspective of collaborative filtering, and similar projects information is consulted during the recommendation process. Thung et al. unify the historical feature requests and API document information to recommend API methods~\cite{Thung2013Automatic}. Yuan et al.~\cite{yuan2019api} combine code parsing and text processing on Android tutorials and SDK documents to recommend functional APIs in Android. Ponzanelli et al. propose a holistic recommendation system Libra, which integrates the IDE and the web browser~\cite{ponz17}. Libra could provide more personalized recommendations since it records developers' navigation history and other contextual information.
\textcolor{black}{FEMIR~\cite{asaduzzaman2017recommending} collects open source software projects hosted on Github to obtain code examples. With static analysis techniques, 
FEMIR mines and organizes the usage patterns for framework extensions, recommending a set of code examples to illustrate all of its relevant extension patterns given user requests.}
\textcolor{black}{Similarly, CSCC~\cite{6976073} leverages code examples collected from software repositories to extract method contexts and use similarity scores to recommend code completion.}
\smallskip
\noindent \textbf{Generation based code recommendation/completion.}
Another important thread mostly bases their methodology on deep learning related techniques~\cite{lecun2015deep}. White et al. empirically demonstrate that a relatively simple RNN model can outperform n-gram models at certain software engineering tasks, such as code suggestion~\cite{white2015toward}. Gu et al. ~\cite{gu2016deep} propose DeepAPI, which adapted a neural language model to encode the words of the query and associated API sequences. By training the model with a large corpus of annotated API from GitHub, DeepAPI could generate API usage sequences for the query. In their subsequent work~\cite{gu2018deep}, a deep neural network model, i.e., CODEnn, was proposed to bridge the lexical gap between queries and source code. It can generate a unified vector representation for both code and descriptions. Liu et al.~\cite{liu2019} leverage autoencoder 
for Android API recommendation tasks. Raychev et al.~\cite{raychev2014code} combine 3-gram and RNN models to synthesize a code snippet, which can complete method invocation and invocation parameters.  Despite that such thread of research mainly generates target code entities, they could still be plugged into our framework, as long as an initial API recommendation list could be produced.


\smallskip
\noindent \textbf{Ranking recommendation results.}
Apart from different approaches towards code recommendation, a few initiatives have focused on applying machine learning based techniques to rank the recommendation candidates. Thung et al. \cite{thung2017webapirec} propose an automated approach, namely WebAPIRec, which can convert web API recommendation into a personalized ranking task based on the API usage historical data. WebAPIRec can learn a model which minimizes errors of Web APIs ordering. Different from our work, WebAPIRec does not utilize  feedback information during recommendation. Wang et al.~\cite{wang2014active} incorporate the feedback into the code search process and propose an active code search approach, which builds the refinement technique on top of the tool  Portfolio~\cite{mcmillan2011portfolio}. For a given query, it first obtains the search result of Portfolio. User opinions for each fragment on the list is collected as feedback and the query representation is expanded. The list is then re-ranked based on the similarity score between the current and the expanded queries. Though the work leverages the feedback information as ours, it addresses the code fragment search problem. Besides, the LTR technique is not utilized. Liu et al.~\cite{liu2018effective} propose a ranking-based discriminative approach, RecRank, to optimize the top-1 recommendation on top of APIREC. Specially, it uses the usage path based features to rank the recommendation list generated by APIREC~\cite{nguyen2016api}. In contrast, our approach does not bind with any particular component recommendation method. In addition, RecRank does not consider the feedback information either. Niu et al.~\cite{niu2017learning} apply the LTR technique to recommend code examples given a query. A pair-wise LTR algorithm is employed to train a ranking schema, which can be  used for new queries later. They address a different recommendation problem, through LTR techniques as well. Moreover, feedback information is also neglected in their approach.

\section{Conclusion}\label{conclusion}
In this paper, we propose BRAID, a novel framework to boost the performance of query-based API recommendation systems. BRAID takes a user query and the result of an existing API recommendation as input. It adopts the user selection history as feedback information and leverages learning-to-rank and active learning techniques to build up a new API recommendation model. With the augmentation of the  feedback information, BRAID performs increasingly better comparing with the baseline API recommenders. The experiments show that BRAID can substantially enhance the effectiveness of state-of-the-art API recommenders. In the future work, we plan to develop a full-fledged tool based on BRAID as a plugin of current mainstream IDEs to better support programming. In addition, we believe  the approach put forward in the current paper actually has broader applicability whereby we plan to extend it to other recommendation scenarios in software engineering.
\section*{Acknowledgements}
This work was partially supported by the National Key R\&D Program of China (No.\ 2018YFB1003902), the National Natural Science Foundation of China (NSFC, No.\ 61972197), the Natural Science Foundation of Jiangsu Province ( No.\ BK20201292), the Collaborative Innovation Center of Novel Software Technology and Industrialization, and the Qing Lan Project. T. Chen is partially supported by Birkbeck BEI School Project (ARTEFACT), NSFC grant (No.\ 61872340), and Guangdong Science and Technology Department grant (No.\ 2018B010107004).

\bibliographystyle{IEEEtran}
\bibliography{draft}




\end{document}